\documentclass[times,twocolumn,final]{elsarticle}

\usepackage{medima}
\usepackage{framed,multirow}
\usepackage{cuted}
\usepackage{caption}
\usepackage{ragged2e}

\usepackage{amssymb}
\usepackage{latexsym}
\usepackage{threeparttable}

\usepackage{url}
\usepackage{color,colortbl,booktabs}
\usepackage[table,x11names,dvipsnames,table]{xcolor}
\definecolor{lightgray}{gray}{0.95}

\usepackage{hyperref}
\hypersetup{
    colorlinks=true,
    linkcolor=blue,
}
\usepackage{rotating}

\usepackage{float,tablefootnote,threeparttable,wrapfig}

\usepackage{booktabs}

\usepackage{lipsum}
\usepackage{color}
\usepackage{diagbox} 
\usepackage{tabu}
\usepackage{arydshln} 
\usepackage{amsmath}
\usepackage{makecell}
\usepackage[ruled]{algorithm}
\usepackage{algpseudocode}

\DeclareMathOperator*{\argmin}{arg\,min}
\newcolumntype{C}[1]{>{\centering\let\newline\\\arraybackslash\hspace{0pt}}m{#1}}
\newcolumntype{L}[1]{>{\raggedright\let\newline\\\arraybackslash\hspace{0pt}}m{#1}}

\newcommand{\revision}[1]{\textcolor{black}{#1}} 
\newcommand{\secRevision}[1]{\textcolor{black}{#1}} 
\newcommand{\final}[1]{\textcolor{black}{#1}} 

\journal{Medical Image Analysis}

\begin{document}

\verso{Y. Bi \textit{et~al.}}

\begin{frontmatter}

\title{
Synomaly Noise and Multi-Stage Diffusion: A Novel Approach for Unsupervised Anomaly Detection in \revision{Medical Images}
}

\author[1,2]{Yuan Bi\corref{cor1}}

\author[1,2]{Lucie Huang\corref{cor1}}
\cortext[cor1]{Authors contributed equally.}

\author[3]{Ricarda Clarenbach}
\author[3]{Reza Ghotbi}

\author[4]{Angelos Karlas}

\author[1]{Nassir Navab}

\author[1]{Zhongliang Jiang}

\address[1]{Computer Aided Medical Procedures, Technical University of Munich, Munich, Germany}
\address[2]{Munich Center of Machine Learning, Munich, Germany}
\address[3]{Clinic for Vascular Surgery, Helios Klinikum M{\"u}nchen West, Munich, Germany}
\address[4]{Department for Vascular and Endovascular Surgery, rechts der Isar University Hospital, Technical University of Munich, Munich, Germany}

\received{XX June 2021}
\finalform{xx Month 2021}
\accepted{xx Month 2021}
\availableonline{xx Month 2021}

\begin{abstract}
\revision{Anomaly detection in medical imaging plays a crucial role in identifying pathological regions across various imaging modalities, such as brain MRI, liver CT, and carotid ultrasound (US). However, training fully supervised segmentation models is often hindered by the scarcity of expert annotations and the complexity of diverse anatomical structures.}
To address these issues, we propose a novel unsupervised anomaly detection framework based on a diffusion model that incorporates a synthetic anomaly (Synomaly) noise function and a multi-stage diffusion process. Synomaly noise introduces synthetic anomalies into healthy images during training, allowing the model to effectively learn anomaly removal. The multi-stage diffusion process is introduced to progressively denoise images, preserving fine details while improving the quality of anomaly-free reconstructions. The generated high-fidelity counterfactual healthy images can further enhance the interpretability of the segmentation models, as well as provide a reliable baseline for evaluating the extent of anomalies and supporting clinical decision-making. Notably, the unsupervised anomaly detection model is trained purely on healthy images, eliminating the need for anomalous training samples and pixel-level annotations. We validate the proposed approach on brain MRI, liver CT datasets, and carotid US. The experimental results demonstrate that the proposed framework outperforms existing state-of-the-art unsupervised anomaly detection methods, achieving performance comparable to fully supervised segmentation models in the US dataset. 
\revision{Ablation studies further highlight the contributions of Synomaly noise and the multi-stage diffusion process in improving anomaly segmentation. These findings underscore the potential of our approach as a robust and annotation-efficient alternative for medical anomaly detection.}
    \textbf{Code:} \url{https://github.com/yuan-12138/Synomaly}.
\end{abstract}

\begin{keyword}
\KWD 
Medical Image Analysis, Unsupervised Anomaly Detection, Diffusion Model, Ultrasound Image Analysis
\end{keyword}

\end{frontmatter}



\begin{figure*}[ht!]
    \centering
    \includegraphics[width=0.95\textwidth]{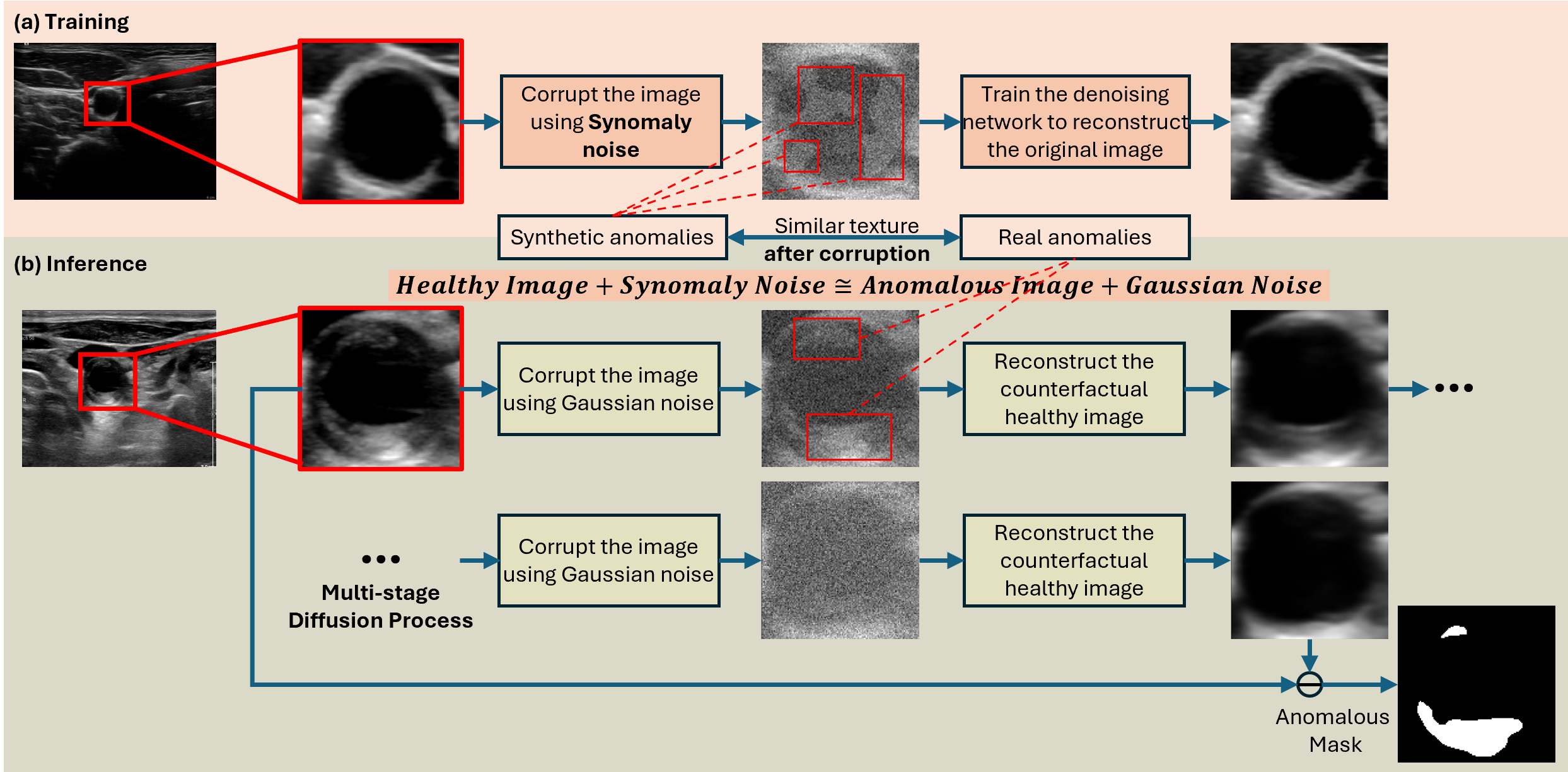}
    \caption{Overview of the proposed approach.(a) illustrates the training process of the proposed unsupervised anomaly detection approach, where Synomaly noise is utilized to corrupt the healthy image. (b) depicts the inference processs, in which Gaussian noise is applied directly to the anomalous images. Since the trained model has already been exposed to similar synthetic anomalies during training, it can efficiently erase the abnormalities. Multi-stage diffusion process is implemented to conserve fine details of the original image, thus increasing the anomaly detection accuracy.}
    \label{fig:overview}
\end{figure*}

\section{Introduction}\label{introduction}

\revision{Anomaly detection in medical imaging is a common task in medical image analysis, where pathological regions are identified for further evaluation and classification. For example, detecting suspicious tumor regions in the brain and liver from magnetic resonance imaging (MRI) or computed tomography (CT) scans, or identifying carotid plaques in ultrasound (US) images, is essential for assessing disease severity and guiding clinical decisions. Traditionally, these tasks have been performed manually by radiologists and sonographers—a process that is time-consuming, labor-intensive, and prone to inter-operator variability.}

\par
\revision{With recent advances in deep learning, automated image analysis has emerged as a powerful tool to assist medical professionals in detecting and evaluating pathologies more efficiently~\citep{jiang2024class, ronneberger2015u, jiang2024intelligent, bi2023mi, huang2025vibnet, jiang2023defcor}. However, the effectiveness and robustness of deep learning models depend heavily on training data quality. In the natural image domain, data collection is relatively straightforward, with images readily available online or captured using consumer-grade cameras. Annotations can be performed by large groups of non-experts, enabling rapid labeling of extensive datasets at a low cost. In contrast, medical imaging data is restricted by privacy regulations and ethical considerations, requiring institutional approval and de-identification~\citep{jiang2023robotic}. Annotation is costly and time-consuming, as it demands specialized expertise from physicians~\citep{bi2024machine}. Additionally, inter-operator variability in labeling can introduce bias, further limiting model generalization. To tackle this challenge, unsupervised anomaly detection approach appears to be an effective solution.}

\par
Similar to how humans learn about diseased tissues, unsupervised anomaly detection method tends to distinguish the anomaly tissue from the normal anatomy by learning from healthy structures~\citep{fernando2021deep}. 
\secRevision{Generative models are commonly utilized to synthesize counterfactual healthy samples from diseased images, enabling anomaly detection without requiring pixel-level annotations \citep{cai2023dual}.} This eliminates the need for extensive manual labeling, which can be challenging due to variations in image quality, contrast, and noise across different imaging modalities.
For example, the high noise ratio and low contrast of US make the labelling process for US sometimes introduces unavoidable ambiguities, thus complicating the fully-supervised training process~\citep{duque2024ultrasound}. However, such drawback does not concern the unsupervised anomaly detection method. Furthermore, it also provides the possibility to detect unseen or rare anomaly cases, which are not well represented in the dataset for fully-supervised models.
\revision{Despite its advantages, the effectiveness of unsupervised anomaly detection largely depends on the generative model’s ability to preserve healthy structures and accurately reconstruct pathological regions as their healthy counterparts. This underscores the need for further exploration of advanced generative models to enhance performance.}
\par
Diffusion models have proven effective in modeling complex data distributions, resulting in higher image sample quality compared to generative adversarial networks (GANs) and variational autoencoders (VAEs)~\citep{dhariwal2021diffusion,bond2021deep, medghalchi2024prompt2perturb}. 
These advantages are particularly valuable for unsupervised anomaly segmentation, especially in cases of small datasets, leading to growing interest in applying diffusion models to this application domain~\citep{kazerouni2023diffusion}. \revision{Recent works have also reported several attempts of applying diffusion models for unsupervised anomaly detection in fetal brain US~\citep{olsen2024unsupervised,mykula2024diffusion}, brain MRI~\citep{liang2024itermask2}, and brain CT~\citep{fontanella2024diffusion}.}
The performance of diffusion model is closely linked to the choice of corrupting noise, especially for medical image anomaly detection~\citep{kascenas2023role}. Superior performance over naive pixel-wise Gaussian noise has been reported when applying Simplex noise~\citep{wyatt2022anoddpm} and Coarse noise~\citep{kascenas2022denoising} for anomaly detection of brain MRI.

\par
In order to develop a robust unsupervised anomaly detection model for \revision{medical} images, in this work, we propose Synomaly noise, which stands for synthetic anomaly noise based on statistical disease morphology prior. Different from the aforementioned noise functions, Synomaly noise is specifically designed for unsupervised medical anomaly detection by corrupting healthy images with synthetic anomalies during training. This design of noise can result in a better erasure of anomalies in the images. To further suppress the loss of detailed information during noising process, a multi-stage diffusion process is proposed for inference. The detailed contributions are summarized as follows:
\begin{itemize}

  \item We introduce the Synomaly noise function to enhance the effectiveness of anomaly detection. This noise is applied to healthy training images to simulate the appearance of real anomaly tissues after the noising process. By exposing the model to these synthetic anomalies during training, the effectiveness of generating healthy images from diseased images is significantly improved, tailored to individual diseases.

  \item We propose a multi-stage diffusion process in the inference phase designed to preserve healthy tissue details while precisely isolating and removing anomalous disease tissue.  
  Instead of executing the diffusion process with a large number of noising and denoising steps in one single stage, the generation is divided into successive diffusion stages, each with fewer noising and denoising steps. Through such gradual and controlled transformation, the model can preserve finer details in the reconstructed images, thus increasing the accuracy in anomaly detection tasks.
  
  \item \revision{We conduct extensive validation across multiple medical imaging modalities, including brain MRI, liver CT, and carotid ultrasound (US) datasets. To the best of our knowledge, this is the first work to apply a diffusion model for unsupervised anomaly detection in carotid US images. Our approach achieves superior performance compared to state-of-the-art methods.}
\end{itemize}
The results of the three most common medical image modalities (US, CT, and MRI) demonstrate the potential of the proposed method in wide application. To facilitate the reproducibility of this work, the code will be released on this website upon acceptance.\footnote{Codes: \url{https://github.com/yuan-12138/Synomaly}}

\section{Related Work}
Anomaly segmentation is based on the principle that a network trained solely on normal images will struggle to reconstruct regions containing abnormalities. The model learns to accurately recreate typical patterns of healthy tissue but fails when facing with abnormal areas. During inference, the model attempts to reconstruct the entire input image, but the anomalous regions remain poorly reconstructed due to the model's lack of exposure to such data. By subtracting the reconstructed image from the original input image, the differences—representing the anomalies—can be highlighted, which enables the pixel-wise identification of abnormal regions~\citep{baur2021autoencoders}.
From a methodological perspective, anomaly segmentation approaches can also be grouped into two main categories: generative autoencoders and denoising models. The following sections will provide an overview of related works in these two categories.

\subsection{Unsupervised Anomaly Detection Using Autoencoders}

Autoencoders (AEs) have been widely used for unsupervised anomaly detection by learning latent representations of healthy anatomy and identifying deviations as anomalies. \secRevision{The reconstructed outputs typically preserve normal regions while modifying anomalous ones, aiding detection and segmentation. Various AE variants have been applied across modalities~\citep{akcay2018ganomaly, mao2020abnormality, siddiquee2024brainomaly}.}
For example, \cite{uzunova2019unsupervised} used a conditional VAE for brain MRI segmentation, and \cite{pawlowski2018unsupervised} applied a Bayesian convolutional AE to CT. \cite{li2023self} proposed SSL-AnoVAE by combining VAE with self-supervised learning for retinal images. Instead of standard reconstruction, \cite{zimmerer2018context} adopted a context-encoding VAE trained via inpainting. To improve AE capacity for modeling complex distributions, adversarial methods have been introduced. \cite{baur2019deep} combined VAE with GAN and proposed AnoVAEGAN for brain MRI anomaly detection.

\par
Apart from reconstruction-based approaches, restoration-based methods offer an alternative. Instead of directly passing input images through AE network to generate counterfactual reconstructions, restoration methods focus on finding the closest representation of the abnormal image in the learned latent space of normal images. \secRevision{By projecting the abnormal image onto the normal manifold, the model transforms it into its most similar healthy counterpart \citep{gong2019memorizing}.} Anomalies are then detected by comparing the original input with the restored image.
\cite{schlegl2017unsupervised} proposed AnoGAN, where a GAN is trained on healthy images to model the latent manifold for normal images. During inference, a random latent point is sampled and iteratively moved towards the closest counterpart on the normal manifold using a gradient decent method. In an extended work, \cite{schlegl2019f} implemented an encoder network to replace the time-consuming iterative restoration process. \cite{chen2020unsupervised} trained a VAE to model the normative data distribution. The restoration process of anomaly samples is performed through an iterative gradient ascent method by Maximum-A-Posteriori estimation. 

\subsection{\secRevision{Unsupervised Anomaly Detection Using Synthetic Anomalies}}
\secRevision{Another mainstream direction in unsupervised anomaly detection is the synthetic anomaly approach, where the goal is to generate anomalous-looking images from normal ones to train a discriminative model. For example, CutPaste \citep{li2021cutpaste} creates pseudo-anomalies by cutting and pasting patches within a normal image to simulate realistic-looking defects. These synthesized images are then used to train a model to directly predict anomaly maps or scores. To improve realism, Natural Synthetic Anomalies (NSA) \citep{schluter2022natural} employ Poisson image editing to seamlessly blend scaled patches from different images, enhancing the variability and natural appearance of the simulated anomalies. 
Similarly, in the medical domain, \cite{tan2021detecting} applied this paradigm to chest X-ray and US images using Poisson image interpolation, generating anomalies that mimic real pathological structures.
However, these methods do not produce counterfactual reconstructions and are dependent on the fidelity and realism of the generated pseudo-anomalies \citep{tan2020detecting}. If the true anomalies have highly out-of-distribution appearances, these approaches may struggle to generalize. Fundamentally, such methods attempt to model what anomalies might look like, rather than learning what healthy tissue should look like, as is done in reconstruction-based methods.}

\subsection{Unsupervised Anomaly Detection Using Denoising Models}

Recent advances in diffusion models have shown strong generative performance across medical imaging modalities~\citep{dominguez2024diffusion,xu2024stereodiffusion,du2023arsdm,jiang2024self}, which has also benefited unsupervised anomaly detection by enabling more accurate identification of abnormalities~\citep{kazerouni2023diffusion, li2023fast}.
\cite{wyatt2022anoddpm} introduced AnoDDPM, a DDPM-based method for generating counterfactual images using a partial diffusion process to preserve fine image details and improve inference speed. They also adopted Simplex noise over Gaussian for better results on brain MRI.
To avoid accumulated errors and improve sampling efficiency, \cite{pinaya2022fast} proposed a latent diffusion model, operating in compressed latent space rather than image space.
\cite{wolleb2022diffusion} used a Denoising Diffusion Implicit Model, enabling faster inference. Their model was trained on both healthy and diseased images, with generation guided by a pretrained classifier.
\cite{bercea2024diffusion} proposed THOR, which incorporates anomaly maps during inference to focus on suspicious regions while preserving healthy structures.
\cite{fontanella2024diffusion} combined DDPM with ACAT~\citep{fontanella2023acat}, using saliency maps to isolate and process potential anomalies separately, improving anomaly erasure in brain MRI and CT.
To address the high computational cost for high-resolution images like MRI, \cite{behrendt2024patched} proposed a patch-based diffusion model that reconstructs images patchwise, reducing complexity.

\par
Given the nature of diffusion process, the corruption of input images will inevitably lead to the loss of details of the original images. However, too little corruption of the input images may also lead to unsatisfactory anomaly erasure performance. To tackle such trade-off, \cite{kascenas2022denoising} proposed denoising AE (DAE) that applies a single instance of Coarse noise to the input image, rather than sequentially adding noise. This one-time noise corruption alters the original image to resemble an anomalous version. By training the DAE to recover the corrupted images, the model can generate counterfactual normal images from pathological ones during inference. In a follow-up work, \cite{kascenas2023role} highlighted the importance of the noise function in diffusion models and DAEs for anomaly segmentation, showing that the choice of noise significantly impacts the models' ability to distinguish between normal and anomalous regions. Inspired by this insight, \cite{liang2024itermask2} employed a masked noising strategy, where only a selected region of the input image is corrupted. During inference, the DAE is iteratively applied until the generated anomaly mask converges. 

\section{Method}

\secRevision{The performance of diffusion models in unsupervised anomaly detection is closely tied to the design of the noising function. An effective noise function should corrupt the image such that, after noising, anomalous regions become indistinguishable from normal tissue—allowing the model to replace them with healthy structures during denoising. However, classic noise functions struggle with large or highly distinct anomalies, e.g., very bright or dark regions, which often require many noise steps to obscure. This heavy corruption can degrade structural details and lead to blurred, inaccurate reconstructions.} This introduces a precision-recall trade-off: Using fewer noise steps preserves structural integrity and achieves high reconstruction accuracy for healthy tissues, but it increases the risk of failing to detect distinct anomalies, leading to a higher rate of false negatives - characterized by high precision and low recall. On the other hand, using more noise steps enhances the likelihood of detecting anomalies, but it also raises the possibility of false positives, where healthy tissues are incorrectly identified as anomalies - characterized by low precision and high recall.


\par
To tackle these challenges, we propose Synomaly noise. The core idea is to corrupt healthy images by introducing synthetic anomalous regions during training, simulating the appearance of noised anomalous regions, as shown in Fig.~\ref{fig:overview}(a). \secRevision{This enables the model to refine its denoising strategy for reconstructing normal tissue from corrupted regions resembling real corrupted anomalies. During inference, standard Gaussian noise is applied, making abnormal areas resemble the synthetic anomalies seen during training. This similarity allows the model to efficiently generate counterfactual healthy tissue and replace anomalies during denoising.} \secRevision{Importantly, the goal of Synomaly noise is not to produce anatomically realistic lesions, but to create synthetic anomalies that mimic the visual appearance of real anomalies after noise corruption.}


\par
To further improve the conservation of fine details during diffusion process, we employ an iterative anomaly erasure approach, as shown in Fig.~\ref{fig:overview}(b). 
In this method, images undergo multiple rounds of the diffusion process, with anomalies being progressively removed in each iteration until the anomaly map converges.
This iterative process is especially beneficial in cases where anomalies are large, as larger noising steps are typically required to fully obscure the abnormal regions. By splitting these larger noising steps into smaller, incremental steps, the model reduces the risk of losing fine details that often occurs when excessive noise is applied at once. 

\subsection{Diffusion Models}
The diffusion model consists of two parts: a forward (noising) process and a reverse (denoising) process. In the forward process, noise is progressively added to a data sample over a sequence of noise steps, gradually corrupting the data and transforming it into a noise-like distribution. The corrupted image can be described as:
\begin{equation} \label{eq:one-step2}
x_t=\sqrt{\bar{\alpha}_t}x_0+\sqrt{1-\bar{\alpha}_t}\epsilon,
\end{equation}
where $x_0$ represents the original image, $x_t$ is the corrupted image at step $t$, $\epsilon$ depicts the noise function, while $\bar{\alpha}_t$ acts as a schedule, controlling the strength of the corruption at each step $t$. Its value starts from $\bar{\alpha}_0=1$ and ends with $\bar{\alpha}_T=0$. $T$ represents the total noising steps, which we set as $1000$ in our setup. This schedule can be a linear schedule as proposed by \cite{ho2020denoising} or a cosine schedule as introduced by \cite{nichol2021improved}.

\par
The reverse diffusion process aims to invert the forward process by gradually removing the added noise from the data, reconstructing the original data from the noisy state. It can be achieved by using a denoising network with trainable parameter $\theta$, normally a U-shape architecture~\citep{ronneberger2015u}. \cite{ho2020denoising} found that instead of directly predicting the denoised image at each step, predicting the noise term ($\epsilon$) works the best among several possible objectives. Hence, the optimization function is defined as:
\begin{equation} \label{eq:L-simple}
\theta^* = \argmin_{\theta} = \mathbb {E}_{t ,x_0,\epsilon} \left[||\epsilon -\epsilon _\theta (x_t,t)||^2\right]. 
\end{equation}
After the training, given a corrupted image $x_t$ or a complete noise $x_T$ we can generate a novel image by iteratively performing the denoising sampling based on DDPM~\citep{ho2020denoising}:
\begin{equation}
\begin{aligned}
x_{t-1} &=\sqrt{\bar{\alpha}_{t-1}}(\frac{x_t-\sqrt{1-\bar{\alpha}_{t}}\epsilon _\theta (x_t, t)}{\sqrt{\bar{\alpha }_t}}) \\
&+ \sqrt{1-\bar{\alpha }_{t-1}-\sigma ^{2}_t} \epsilon _\theta(x_t,t) + \sigma _t\epsilon 
\end{aligned}
\end{equation}
with \(\sigma _t = \sqrt{(1-\bar{\alpha }_{t-1})/(1-\bar{\alpha }_t)}\sqrt{1-\bar{\alpha }_t / \bar{\alpha }_{t-1}}.\)

\subsection{Synomaly Noise}\label{sec:synomaly}
Synomaly noise stands for \textbf{syn}thetic an\textbf{omaly} noise. Instead of merely adding noise until anomalies blend into healthy tissues, Synomaly strategically integrates synthetic anomalies into the noise model during training. 
\secRevision{The core idea is to train the diffusion model to remove synthetic anomalies embedded in healthy images that resemble real pathological patterns, thereby enhancing its ability to detect and remove true anomalies at inference.}

\begin{figure}[ht!]
    \centering
    \includegraphics[width=0.44\textwidth]{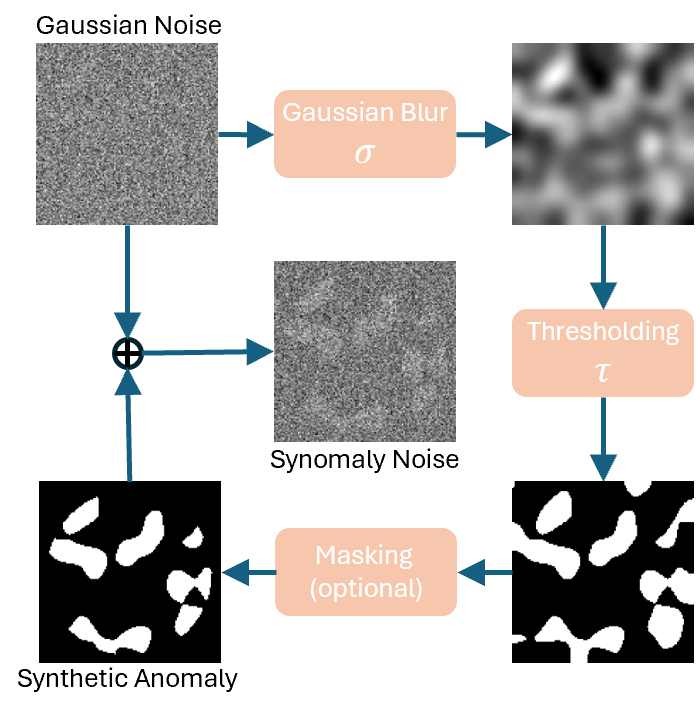}
    \caption{Generation of Synomaly Noise}
    \label{fig:synomaly_generation}
\end{figure}

\par
Fig.~\ref{fig:synomaly_generation} illustrates the process of Synomaly noise generation. In the first step, a pure Gaussian noise is created, matching the dimensions of the input image. This Gaussian noise will also serve as the Gaussian background noise of Synomaly noise. 
Next, the shapes of the synthetic anomalies are generated by applying Gaussian blurring to the initial noise, followed by thresholding. Here, the variance of Gaussian blurring ($\sigma$) as well as the threshold value ($\tau$) are predefined as model training parameters and control the size and number of synthetic anomalies. Specifically, increasing the blurring variance ($\sigma$) results in larger fragments of anomalies, while higher threshold values ($\tau$) reduce the number and size of anomalies by excluding lower Gaussian values. \revision{To further refine the simulation of real anomalies, an optional mask can be applied to ensure that synthetic anomalies are placed within relevant anatomical regions rather than in background areas. This mask is particularly useful for imaging modalities where the foreground and background are not clearly separable or where anomalies are expected to occur within specific structures.}

\par
Once the synthetic anomaly shapes are defined, synthetic anomalies are added to the Gaussian background noise at their respective locations. A predefined anomaly direction $d\in \left\{ -1,1 \right\}$ determines whether the anomalies typically appear brighter or darker than the surrounding healthy tissue and a predefined intensity offset ($i$) dictates the distinctiveness of the synthetic anomalies. Additionally, a random value ($v$) between 0 and 1 is added to the intensity of these synthetic anomalies to enhance robustness against variations in actual anomaly brightness.

\par
In summary, the tunable parameters include blurring variance $\sigma$, threshold value $\tau$, anomaly direction $d$ and intensity offset $i$. Guided by these parameters to reflect realistic medical conditions, the resulting synthetic anomalies vary randomly in number, shape, size, intensity and location, as they are all derived from pure Gaussian noise.

\begin{figure}[ht!]
    \centering
    \includegraphics[width=0.48\textwidth]{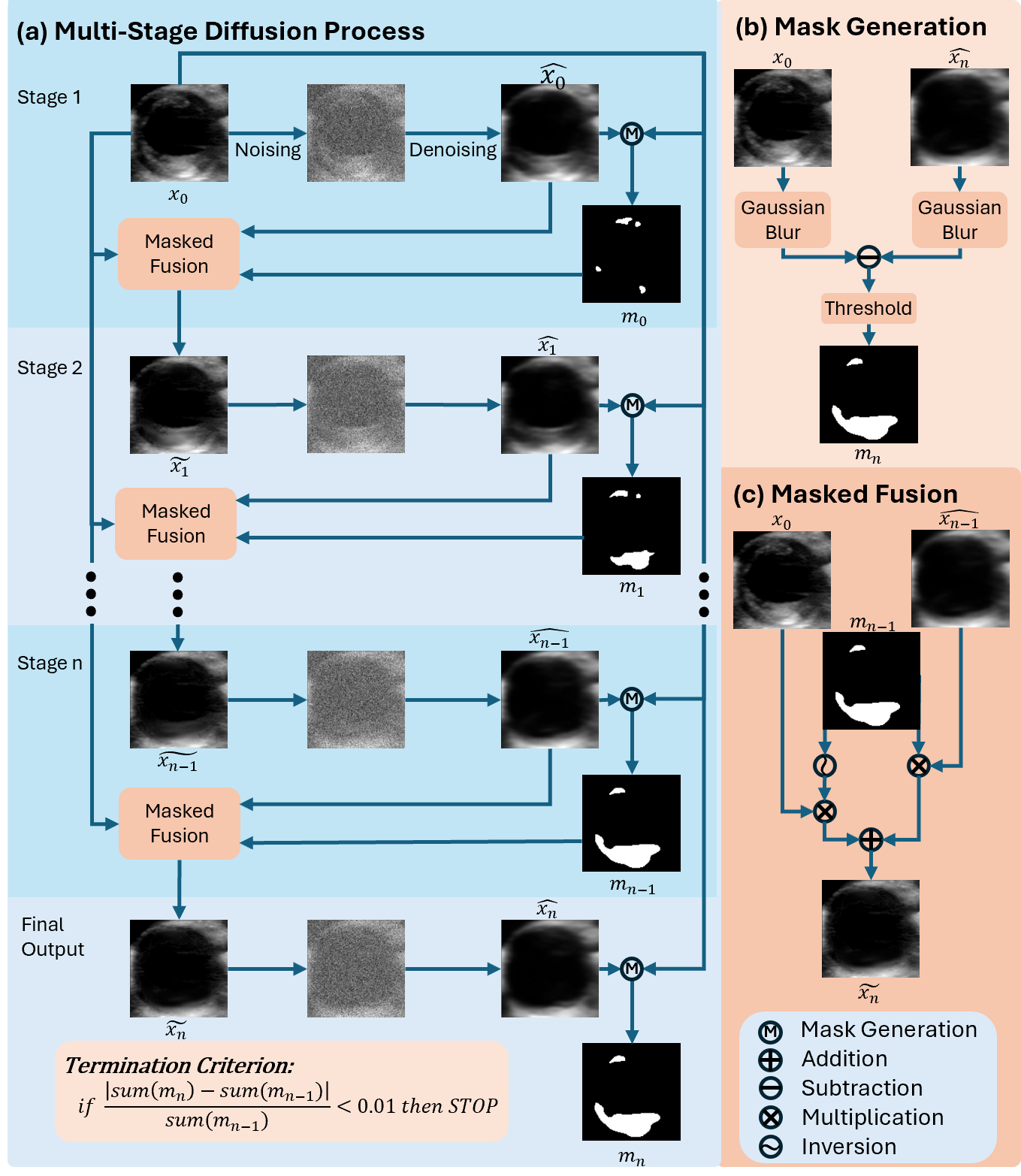}
    \caption{\secRevision{Multi-stage diffusion process with masked fusion.}}
    \label{fig:multi_stage}
\end{figure}

\subsection{Multi-stage Diffusion Process at Inference}
Most of the existing works on diffusion models for anomaly detection typically suggest performing the diffusion process only once during inference, which is termed as ``single-stage'' approach. 
\secRevision{However, this is often inadequate for anomalies with distinct appearance. A single pass may not fully remove such anomalies, especially when many noise steps are needed. While increasing noise steps can improve anomaly removal, it also degrades structural detail, resulting in blurred and inaccurate reconstructions.}

\par
Therefore, we propose a ``multi-stage diffusion process''. Instead of applying many noise steps in a single inference stage, the multi-stage approach repeatedly applies a few noise steps followed by denoising across multiple inference stages. \secRevision{As depicted in Fig.~\ref{fig:multi_stage}(a), starting with the original input image $x_0$, the process begins by adding noise for a predefined number of noise steps, followed by denoising. Based on the denoised image $\widehat{x_n}$, the anomaly mask $m_n$ is generated (Fig. \ref{fig:multi_stage}(b)) by thresholding the difference between blurred versions of the denoised and the original image. Using this mask, masked fusion (Fig. \ref{fig:multi_stage}(c)) combines anomalous regions from $\widehat{x_n}$ with the background from the original image $x_0$, producing a new input $\widetilde{x_n}$ for the next stage.}
\revision{This process iterates until the sum of detected anomaly pixels stabilizes, meaning that the relative change between consecutive anomaly masks falls below a threshold of 0.01. As a result, the number of diffusion stages at inference is not a manually set hyperparameter but is automatically determined based on the convergence of the anomaly mask. To prevent excessive computations, a maximum of five iterations is imposed, as further iterations typically yield negligible improvements.}

\begin{algorithm}[!h]
\caption{Multi-stage Diffusion Process at Inference}
\label{alg:multi_stage_inference}
\begin{algorithmic}

\State \secRevision{Input: input image \(x_0\)}
\State \secRevision{Output: anomaly mask \(m_n\)}

\State \secRevision{$n = 0$}
\State \secRevision{$\widetilde{x_0} = x_0$}
\While{\secRevision{$\frac{|sum(m_{n-1})-sum(m_n)|}{sum(m_{n-1})}>0.01$}}
    \State \secRevision{Apply noising to $\widetilde{x_n}$: $\widetilde{x_{n,t}} \gets \text{noise}(\widetilde{x_n}, t)$}
    \State \secRevision{Apply denoising to $\widetilde{x_{n,t}}$: $\widehat{x_n} = \text{denoise}(\widetilde{x_{n,t}})$}
    \State \secRevision{Calculate anomaly mask: }
    \State \secRevision{~~~~~~$m_n = threshold(|blur(x_0)- blur(\widehat{x_n})|)$}
    \State \secRevision{Apply masked fusion: $\widetilde{x_{n}} = m_n \cdot \widehat{x_{n}} + (1-m_n) \cdot x_0$}
    \State \secRevision{$n = n+1$}
\EndWhile

\State \Return \secRevision{$m_n$}

\end{algorithmic}
\end{algorithm}

This approach offers several benefits. \secRevision{By decomposing a large noise step into multiple smaller steps across stages, the structural integrity of the image is better preserved, leading to more accurate reconstructions.} Additionally, performing multiple inference stages with sequential denoising allows even distinct anomalies to be progressively removed and detected more effectively. \secRevision{Since different noise steps model different image features \citep{ho2020denoising} and the goal of medical anomaly detection is to focus on finer details without excessively corrupting the entire image, fewer noise steps are particularly relevant for preserving details from non-anomalous regions.}

\par
To further reduce the false positive rate in the generated anomaly mask, we propose a ``masked fusion'' strategy, as shown in Fig.~\ref{fig:multi_stage}(b). At each diffusion stage, rather than directly passing the generated image to the next cycle of noising and denoising, we mask the image by the anomaly mask produced at that stage and fuse it with the original input image. This approach preserves the ``normal'' regions from the input image, while maintaining the reduced anomalies at each stage. 
As a result, the quality of the generated healthy image is enhanced, effectively reducing the false positive detection rate. 
The detailed process is summarized in Algorithm~\ref{alg:multi_stage_inference}.

\section{\secRevision{Experiments and Discussions}}
In this section, we evaluate the performance of the proposed model using various datasets. Since the US carotid plaque dataset is collected in-house, we additionally present evaluation results on publicly available MRI and CT datasets to demonstrate the superiority and effectiveness of the proposed method. 

\subsection{Datasets}
\secRevision{We evaluate the proposed method on three representative datasets spanning different modalities and anatomical regions: Brain MRI (BraTS 2023)~\citep{baid2021rsna, bakas2017advancing, menze2014multimodal}, Liver CT (LiTS) \citep{bilic2023liver}, and Carotid US. The datasets comprise both healthy and anomalous cases, with Brain MRI containing 3743 healthy training slices, 1000 anomalous test slices, and 416 healthy test slices; Liver CT including 5820 healthy training images, 1000 anomalous test images, and 647 healthy test images; and Carotid US consisting of 7306 healthy training images, 545 anomalous test images, and 833 healthy test images. Additional details on data acquisition, preprocessing, and sample images are provided in Appendix A.}

\begin{table}[ht!]
\caption{Best hyperparameter combinations of the selected methods for different anomaly segmentation tasks}\label{tab_hyperparameter}
\resizebox{0.48\textwidth}{!}{
\begin{threeparttable}
  \begin{tabular}{l ccc|ccc|ccc}
   \multirow{2}{*}{Method}&\multicolumn{3}{c|}{Noise steps ($\mathcal{T}$)}&\multicolumn{3}{c|}{Kernel size ($n$)}&\multicolumn{3}{c}{Threshold ($Th$)}\\
     & MRI & CT & US & MRI & CT & US & MRI & CT & US \\ \hline
    VAE            & $-$ & $-$ & $-$     & 13 & 13 & 13      & 0.2  & 0.35  & 0.3            \\ 
    DDPM+Gaussian  & 300 & 300 & 800     & 15 & 15 & 15      & 0.15 & 0.2  & 0.3           \\
    DDPM+Corase    & 700 & 800 & 800     & 1  & 5  & 21      & 0.15 & 0.1  & 0.1           \\
    DDPM+Pyramid   & 350 & 250 & 800     & 13 & 13 & 5       & 0.2  & 0.2  & 0.35           \\
    DDPM+Simplex   & 200 & 200 & 700     & 1  & 1  & 5       & 0.2  & 0.25 & 0.45           \\
    AutoDDPM       & 200 & 400 & 700     & 9  & 3  & 13      & 0.1  & 0.2  & 0.1            \\
    DAE            & $-$ & $-$ & $-$     & 11 & 9  & 3       & 0.1  & 0.25 & 0.1           \\ \hline
    DDPM+Synomaly+Single-stage  & 200 & 200 & 400     & 15 & 11 & 11      & 0.15 & 0.25 & 0.25           \\
    DDPM+Synomaly+Multi-stage  & $150^*$ & $150^*$ & $250^*$    & 15 & 5 & 15    & 0.2 & 0.2 & 0.3  \\ \hline
  \end{tabular}
  \begin{tablenotes}
    \item[*] Noise steps at each stage.
  \end{tablenotes}
\end{threeparttable}
  }
\end{table}

\subsection{Inference Hyperparameters}
The number of inference noise steps $\mathcal{T}$ plays a significant role in balancing accurate image reconstruction and effective anomaly detection. As discussed earlier, methods like optimizing the noise function can be employed to address this trade-off, but it remains essential to find the optimal balance between anomaly removal and image preservation for specific tasks. Therefore, it is a critical hyperparameter dominating the anomaly detection results.

\par
\secRevision{Directly computing the pixel-wise difference between the original and generated images is not always the most suitable approach due to potential imperfections in pixel-wise image reconstruction.} Moreover, in the context of preventive medical anomaly detection, the focus lays on identifying regions of potential anomalies rather than individual pixels, so that medical professionals can further evaluate these areas. \secRevision{Therefore, Gaussian blurring ($G_n$) is applied to both the input and generated images before calculating their differences.} The size of the Gaussian kernel is denoted as $n$, denoting a second hyperparameter that influences the anomaly segmentation performance.

\par
In addition to the inference noise steps $\mathcal{T}$ and Gaussian kernel size $n$, a third crucial parameter during inference is the anomaly threshold $Th$. \secRevision{After the difference map is created by calculating the pixel-wise residuals $|x \ast G_n -\hat{x} \ast G_n|$ between the blurred original image $x$ and blurred generated image $\hat{x}$, anomalies are identified by applying this anomaly threshold to the difference map,} flagging regions with differences above this threshold as anomalies. 

\par
In summary, the inference noise steps for partial diffusion ($\mathcal{T}$), the Gaussian kernel size for blurring ($n$), and the anomaly threshold for anomaly selection ($Th$) \revision{are the three primary hyperparameters considered during inference.} 
\revision{However, it is notable that the Gaussian kernel size and threshold value do not impact the quality of the generated counterfactual image. These parameters are used solely in the post-processing step to refine the anomaly mask by controlling the level of smoothing and the sensitivity of anomaly detection. In contrast, the number of noise steps plays a crucial role in determining the balance between preserving fine details and effectively removing anomalies during counterfactual image reconstruction. To ensure a fair comparison between different models, we determine the optimal inference hyperparameters for each approach, representing their best possible performance. As shown in Tab.~\ref{tab_hyperparameter}, the proposed method (DDPM+Synomaly+Multi-stage) shows minimal variation in noise steps across datasets, differing by only 100 steps, demonstrating its robustness to hyperparameter tuning.}

\subsection{Training Details}
The performance of the proposed anomaly detection method is compared with \secRevision{eight state-of-the-art (SOTA) methods}: classic VAE~\citep{kingma2013auto}, \secRevision{f-AnoGAN~\citep{schlegl2019f}, NSA~\citep{schluter2022natural}}, DDPM using Gaussian noise function~\citep{ho2020denoising}, DDPM using Coarse noise function~\citep{kascenas2023role}, DDPM using Pyramid noise function~\citep{frotscher2023unsupervised}, \secRevision{DDPM using Simplex noise function~\citep{wyatt2022anoddpm} (also known as AnoDDPM)}, AutoDDPM~\citep{bercea2023mask}, and DAE~\citep{kascenas2022denoising}. Fully supervised UNets~\citep{ronneberger2015u} were also trained for each dataset as a reference. Half of the anomalous data of each dataset is selected as UNet training data.

\begin{table*}[ht!]

\caption{Performance comparison of the proposed approach with various unsupervised anomaly detection methods. The best results are highlighted in bold, and the second-best results are underlined. For reference, the results of the fully supervised UNet are listed at the bottom.}\label{tab_results}

\resizebox{\textwidth}{!}{
\begin{threeparttable}
  \begin{tabular}{l ccc|ccc|ccc}
    \toprule
    \multirow{2}{*}{Method}&\multicolumn{3}{c}{Brain MRI BraTS23}&\multicolumn{3}{c}{Liver CT LiTS}&\multicolumn{3}{c}{Carotid Plaque US}\\
    \cline{2-10}
     & DSC & Precision & Recall & DSC & Precision & Recall & DSC & Precision & Recall \\
    \hline
    VAE~\citep{ronneberger2015u}                   & $0.490\pm0.226$          & $0.514\pm0.274$          & $0.532\pm0.245$          & $0.131\pm0.108$          & $0.085\pm0.076$          & $0.362\pm0.238$          & $\underline{0.491\pm0.140}$ & $\underline{0.558\pm0.167}$ & $0.486\pm0.184$\\
    \secRevision{f-AnoGAN~\citep{schlegl2019f}}                  & \secRevision{$0.592\pm0.242$}          & \secRevision{$0.609\pm0.277$}          & \secRevision{$0.634\pm0.251$}          & \secRevision{$0.081\pm0.041$}          & \secRevision{$0.043\pm0.023$}          & \secRevision{$\mathbf{0.735\pm0.085}$}          & \secRevision{$0.372\pm0.145$} & $0.340\pm0.156$ & \secRevision{$0.468\pm0.191$}\\
    \secRevision{NSA~\citep{schluter2022natural}}                & \secRevision{$0.519\pm0.156$}          & \secRevision{$0.370\pm0.135$}          & \secRevision{$\mathbf{0.951\pm0.158}$}          & \secRevision{$0.121\pm0.206$}          & \secRevision{$0.182\pm0.296$}          & \secRevision{$0.137\pm0.261$}         & \secRevision{$0.437\pm0.172$} & \secRevision{$0.371\pm0.174$} & \secRevision{$\underline{0.633\pm0.281}$}\\
    DDPM+Gaussian~\citep{ho2020denoising}          & $0.613\pm0.205$          & $0.630\pm0.229$          & $0.638\pm0.228$          & $0.267\pm0.168$          & $0.230\pm0.172$          & $0.383\pm0.227$          & $0.347\pm0.127$          & $0.299\pm0.130$          & $0.475\pm0.188$\\
    DDPM+Corase~\citep{kascenas2023role}           & $0.592\pm0.262$          & $0.637\pm0.299$          & $0.585\pm0.280$          & $0.402\pm0.247$          & $0.403\pm0.260$          & $0.469\pm0.304$          & $0.230\pm0.100$          & $0.150\pm0.074$          & $0.566\pm0.151$\\
    DDPM+Pyramid~\citep{frotscher2023unsupervised} & $0.620\pm0.236$          & $\underline{0.743\pm0.266}$ & $0.566\pm0.241$          & $0.283\pm0.201$          & $0.301\pm0.236$          & $0.307\pm0.220$          & $0.344\pm0.116$          & $0.298\pm0.115$          & $0.448\pm0.171$\\
    DDPM+Simplex~\citep{wyatt2022anoddpm}          & $0.613\pm0.257$          & $0.689\pm0.302$          & $0.595\pm0.255$          & $\underline{0.537\pm0.246}$ & $\underline{0.573\pm0.284}$ & $0.581\pm0.276$ & $0.413\pm0.162$          & $0.354\pm0.168$          & $0.595\pm0.262$\\
    AutoDDPM~\citep{bercea2023mask}                & $\underline{0.629\pm0.202}$ & $0.627\pm0.244$          & $0.702\pm0.209$ & $0.221\pm0.143$          & $0.157\pm0.113$          & $0.444\pm0.252$          & $0.268\pm0.142$          & $0.285\pm0.163$          & $0.267\pm0.148$\\
    DAE~\citep{kascenas2022denoising}              & $0.553\pm0.264$          & $0.544\pm0.344$          & $0.664\pm0.265$          & $0.445\pm0.266$          & $0.467\pm0.311$          & $0.524\pm0.297$          & $0.213\pm0.082$          & $0.139\pm0.065$          & $0.537\pm0.146$\\ 
    DDPM+Synomaly+Multi-stage                      & $\mathbf{0.739\pm0.234}$ & $\mathbf{0.786\pm0.247}$ & $\underline{0.741\pm0.259}$          & $\mathbf{0.577\pm0.303}$ & $\mathbf{0.622\pm0.302}$ & $\underline{0.600\pm0.312}$ & $\mathbf{0.731\pm0.117}$ & $\mathbf{0.748\pm0.153}$ & $\mathbf{0.772\pm0.180}$\\
    \hline
    UNet~\citep{ronneberger2015u}$^\ast$  &$0.828\pm0.223$&$0.850\pm0.237$&$0.850\pm0.217$&$0.849\pm0.127$&$0.861\pm0.133$&$0.854\pm0.143$ &$0.797\pm0.065$&$0.867\pm0.079$&$0.750\pm0.109$ \\ \hline
  \end{tabular}
  \begin{tablenotes}
    \item[*] Fully supervised method.
  \end{tablenotes}
\end{threeparttable}
  }

\end{table*}

\par
\secRevision{The VAE model, f-AnoGAN, NSA model, and DAE model are trained for $2000$ epochs using Adam optimizer with an exponentially decreasing learning rate from $1\times 10^{-4}$ to $1 \times 10^{-5}$ by $0.9$.} All the DDPM models are implemented based on the popular guided diffusion approach by OpenAI \citep{dhariwal2021diffusion}. For the Simplex noise function, the noising parameters were set with an octave of 6, a persistence of 0.8 and a frequency of 64, as recommended by the original authors. For Coarse noise, the noising parameters were configured with a noise resolution of 16 and a noise standard deviation of 0.2, also aligning with the established literature. All the DDPMs are trained using Adam optimizer with a fixed learning rate of $1\times10^{-5}$. The models were trained for 3000 epochs on the brain MRI and liver CT datasets, and for 2000 epochs on the carotid US dataset. All the models were trained using Nvidia A40 GPU. \secRevision{During inference, DDIM~\citep{song2020denoising} is applied to accelerate the inference process.}

\subsection{\revision{Pixel-level Anomaly Detection Performance}}
\revision{In this section, we evaluate the performance of the proposed method on pixel-level anomaly detection, comparing it with SOTA methods. The evaluation is conducted using anomalous images from the test set, where segmentation performance is measured to assess how effectively each method detects and localizes anomalies.} \secRevision{Quantitative results are reported in Tab.~\ref{tab_results}, while the visual results are shown in Fig.~\ref{fig:experiments_results}. Since NSA is not a generative model and only outputs anomaly masks, its results are not shown in Fig.~\ref{fig:experiments_results}.}

\subsubsection{Results on Brain MRI}
For the anomaly detection in brain MRI images, the Synomaly noise function parameters are configured with a blurring variance \(\sigma\) of 3, a threshold value \(\tau\) of 175, an anomaly direction \(d\) of 1 and an intensity offset \(i\) of 0.5. These Synomaly parameters were determined through a combination of known disease morphology and experimental validation.
\revision{No mask is applied when generating Synomaly noise for brain MRI, as the brain occupies the majority of the image and is easily distinguishable from the background, reducing the risk of anomalies being misplaced in irrelevant areas.}
To ensure a fair comparison between methods, the optimal inference hyperparameters for each method were determined through a comprehensive grid search. Tab.~\ref{tab_hyperparameter} presents the best hyperparameter combinations for each method. Notably, for VAE and DAE, since no diffusion process is involved, the ``noise steps'' field is left blank. 
``DDPM+Synomaly+Multi-stage'' depicts the case where the proposed Synomaly noise is used to train the DDPM and the multi-stage diffusion process is utilized during inference. The ``noise steps'' field for ``DDPM+Synomaly+Multi-stage'' indicates the noise steps at each stage of the iterative process.

\par
Using the best parameters identified, each model was tested on 1000 anomalous images. The evaluation metrics include the Dice, Precision, and Recall score, to ensure an objective and comprehensive comparison of different approaches. The anomaly segmentation results are shown in Tab.~\ref{tab_results}. 
\secRevision{It is evident that the proposed method consistently achieves the highest performance in terms of DSC and precision. While NSA yields the highest recall in the brain MRI dataset, its substantially lower precision suggests that it tends to over-segment, reducing its overall effectiveness.}
Fig.~\ref{fig:experiments_results} shows the visual results of the reconstructed images and anomaly segmentation maps using different unsupervised methods. 
DDPMs trained using Coarse noise and Simplex noise functions partially erase the anomalous regions, but distinct borders of the brain tumors remain in the reconstructed images. For AutoDDPM and DAE, the bright anomalous area is slightly dimmed but not completely removed. However, changes in these areas can still be identified with a proper threshold. \secRevision{The reconstruction results of VAE and f-AnoGAN are anomaly-free but appears blurry, losing important brain details.} Meanwhile, DDPMs trained with Gaussian noise and Pyramid noise fully remove the round anomaly in the top area of the image but fail to accurately reconstruct the lateral ventricle. 
In contrast, the image reconstructed using the proposed method shows a more realistic reconstruction, particularly for the lateral ventricle, demonstrating its superior performance in terms of generative fidelity. The proposed approach delivers the best anomaly segmentation mask compared to the other SOTAs.


\subsubsection{Results on Liver CT}

The configuration of Synomaly noise function for liver CT anomaly detection task is set as a blurring variance \(\sigma\) of 3, a threshold value \(\tau\) of 175, an anomaly direction \(d\) of -1 and an intensity offset \(i\) of 1. Note that for this specific use case, the anomaly detection direction was set to -1 instead of 1, since the liver tumors appear darker than the surrounding healthy tissue, requiring an inversion of the detection parameter. \revision{The generated Synomaly noise is masked using the liver segmentation provided in the dataset, ensuring that synthetic anomalies are predominantly placed inside the liver region.}

\par
\begin{figure*}[ht!]
    \centering
    \includegraphics[width=0.94\textwidth]{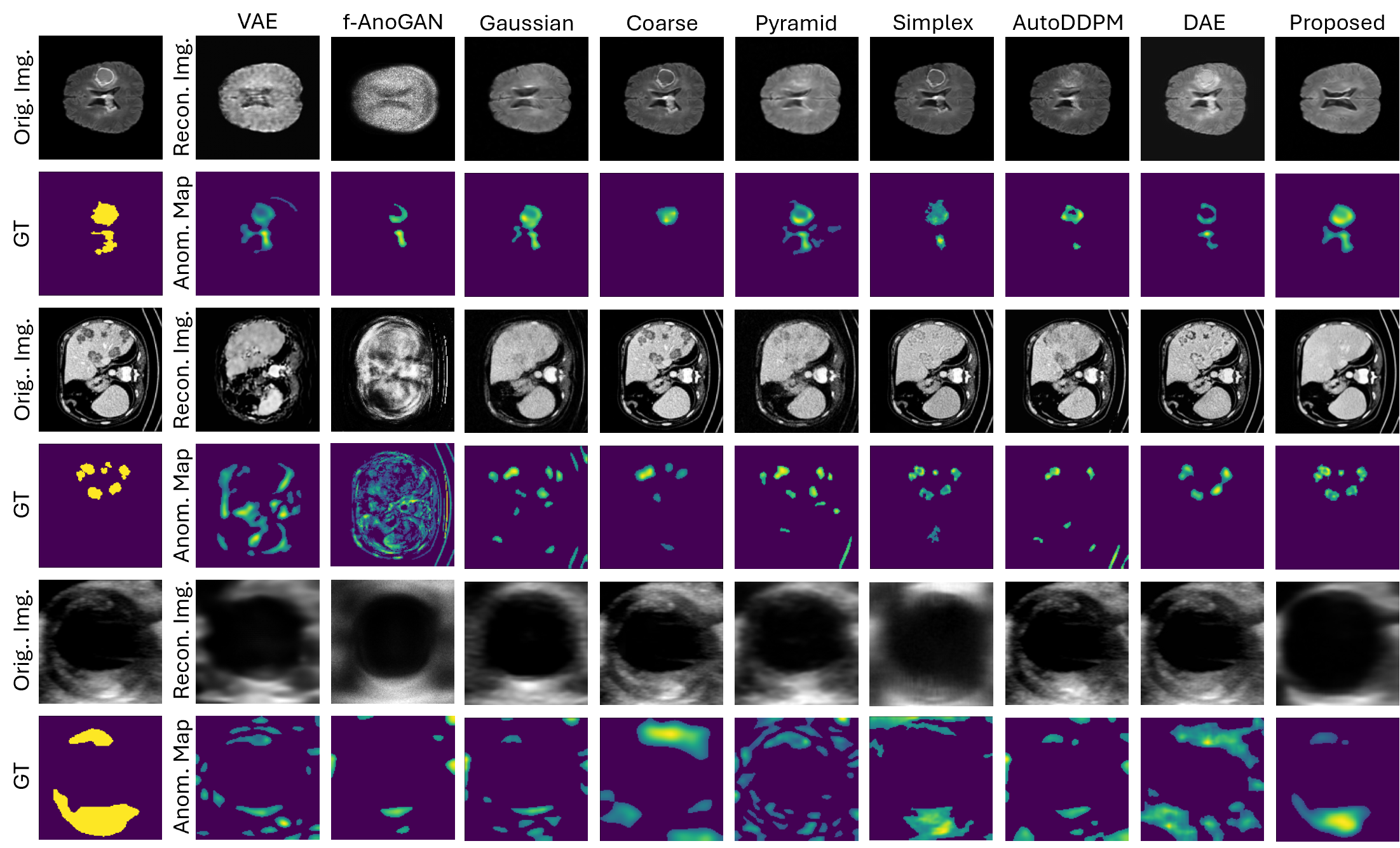}
    \caption{Samples of anomaly segmentation results using different unsupervised methods.}
    \label{fig:experiments_results}
\end{figure*}

The best inference hyperparameters are determined for each method by applying grid search. The anomaly segmentation performance is summarized in Tab.~\ref{tab_results}. \secRevision{In this scenario, the VAE, f-AnoGAN, and NSA completely fail to generate plausible counterfactual samples, as the complexity of CT images exceeds the generative capacity of the VAE model.} \secRevision{As shown in Fig.~\ref{fig:experiments_results}, the generated images of VAE and f-AnoGAN are of low resolution with jagged borders.} The DDPMs trained with Gaussian and Pyramid noise function can effectively erase the abnormalities but they also corrupt other parts of the image, resulting in a high false-positive detection rate. The AutoDDPM and DAE cannot fully remove the anomalies, but with a proper threshold, the anomalous regions can still be highlighted. The DDPM trained with the Simplex noise function demonstrated relatively good generation results, with liver tumors almost fully removed. 
The best performance was observed with Synomaly noise plus the multi-stage diffusion process, which achieved the highest scores in DSC and precision. The visual results also confirm its superior performance, as the anomalies are completely erased without any trace. When comparing the performance of unsupervised anomaly segmentation methods to fully supervised segmentation models, a significant gap remains in this scenario. This can be attributed to the complexity of abdominal CT images, where the generative model must reconstruct not only the liver but also all other organs in the abdomen, making accurate segmentation considerably more challenging.

\subsubsection{Results on Carotid US}

The Synomaly noise function in this scenario was defined with parameters based on known disease morphology and experimental validation, which resulted in a blurring variance \(\sigma\) of 7, a threshold value \(\tau\) of 150, an anomaly direction \(d\) of 1 and an intensity offset \(i\) of 0.5. \revision{A circular mask is applied to restrict synthetic anomalies within the vessel region, ensuring that the generated anomalies align with the expected anatomical distribution. The mask has a diameter of 90\% of the image size and is centered within the image to approximate the vessel’s location.} 
To identify the best inference parameters for each model, a comprehensive grid search was conducted across the parameters noise steps ($\mathcal{T}$), Gaussian kernel size ($n$) and anomaly threshold ($Th$). Tab.~\ref{tab_hyperparameter} presents the optimal parameters for each method.

\par
The anomaly segmentation results are summarized in Tab.~\ref{tab_results}. It is evident that Synomaly noise significantly outperforms all other unsupervised methods. Interestingly, the best performance apart from the proposed approach in this scenario is achieved by the VAE, with an average Dice score of $0.491$. \secRevision{This can be attributed to the fact that, compared to brain MRI and liver CT, the structure of the carotid vessel is relatively simple, allowing the generative models like GANs and VAEs to reconstruct a plausible plaque-free artery from a diseased one.} Additionally, carotid plaques are more distinct and larger in size, which makes removing such anomalies with traditional SOTA noise functions challenging. Large noising steps are required, resulting in severe corruption of the image and significant information loss. Due to the similarity in textures between plaques and surrounding tissues, noising further blurs the distinction between these structures. As observed in the fifth row of Fig.~\ref{fig:experiments_results}, generated vessels often incorrectly interpret the inner border of the original diseased vessel as the outer border, leading to reconstructions of smaller-than-actual arteries.
The DAE fails to effectively remove the plaques partly because the added noise during training does not sufficiently represent the anomalies. Consequently, during inference, the network encounters situations that it was not exposed to during training. In contrast, Synomaly noise effectively identifies and removes the plaques within the artery. 
The image generated using the proposed approach demonstrates the best performance among all methods evaluated. As illustrated in the last subplot of the fifth row of Fig.~\ref{fig:experiments_results}, the plaques are completely removed, and the artery's borders are accurately reconstructed.
\par
Compared to the fully supervised method, the proposed unsupervised approach is only $0.06$ points lower in Dice score and even achieves a higher Recall. Considering the small size of the carotid plaque dataset compared to brain MRI and liver CT, the reduced segmentation performance of fully supervised method can be attributed to the scarcity of diseased data. However, this limitation reflects the typical reality of US imaging, where large diseased datasets are difficult to obtain, while healthy datasets are easily accessible. Notably, the healthy US dataset was collected using a Siemens machine using default ``Carotid'' acquisition parameter setting, whereas the anomalous dataset was recorded using a Philips machine with a sonographer-defined parameter setting. Despite the existing domain gap, the proposed anomaly detection model still demonstrates comparable performance to the fully supervised method, highlighting its robustness and effectiveness in practice.

\par
\revision{We acknowledge that, to reduce complexity, a vessel tracking step was applied to localize the carotid artery region and minimize irrelevant background information. This preprocessing mitigates the high variability of US backgrounds, which can mislead generative models. Unlike CT and MRI, where background structures provide a stable anatomical reference, US backgrounds are inconsistent, making anomaly detection more challenging. Despite this simplification, the task remains difficult—even a fully supervised segmentation model trained on zoomed-in anomalous images achieved only 0.797 Dice, highlighting that anomaly detection remains a challenging task even after vessel localization.}

\par
\revision{To demonstrate that vessel tracking can be achieved with minimal effort, we conducted an additional experiment using Segment Anything Model 2 (SAM2)~\citep{ravi2025sam} for automated carotid vessel tracking before applying the proposed anomaly detection method. 
Only the ground truth vessel mask of the first frame in each US sweep was provided to SAM2 for initialization, after which the model tracked the vessel across subsequent frames, and images were cropped based on the tracking results.
Without any fine-tuning on US data, SAM2 successfully tracked the carotid region. Based on the tracking results, the final anomaly segmentation achieved a Dice score of $0.680\pm0.188$. This suggests that modern large segmentation models can perform robust vessel tracking with minimal manual intervention. However, while vessel tracking benefits from strong shape priors, this may not generalize to more irregular structures like anomalies, where tracking remains challenging. These results highlight the feasibility of automated tracking as a practical preprocessing step for improving anomaly detection in US imaging.}

\begin{figure}[ht!]
    \centering
    \includegraphics[width=0.45\textwidth]{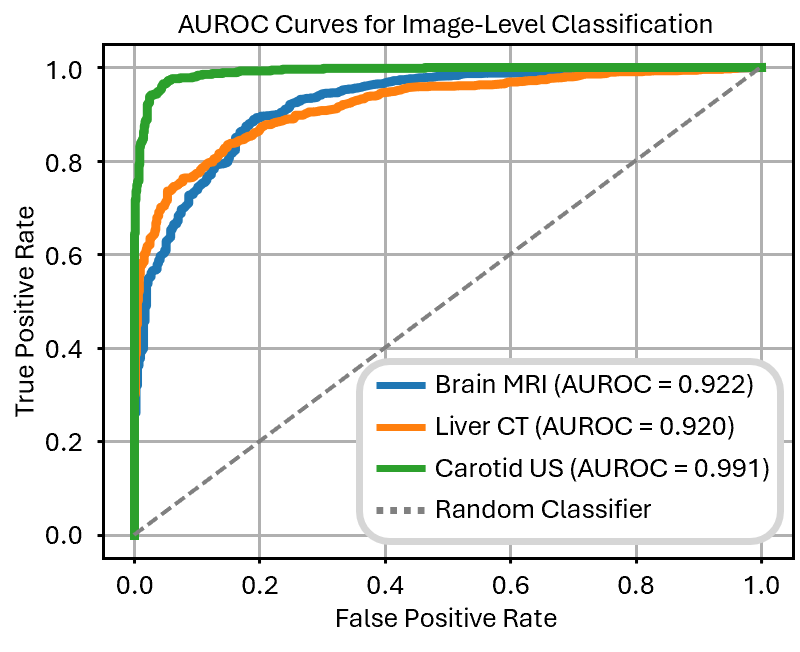}
    \caption{\revision{AUROC curves for image-level classification using the proposed Synomaly noise and multi-stage diffusion across different medical imaging datasets.}}
    \label{fig:auroc_curves}
\end{figure}

\begin{figure*}[ht!]
    \centering
    \includegraphics[width=0.98\textwidth]{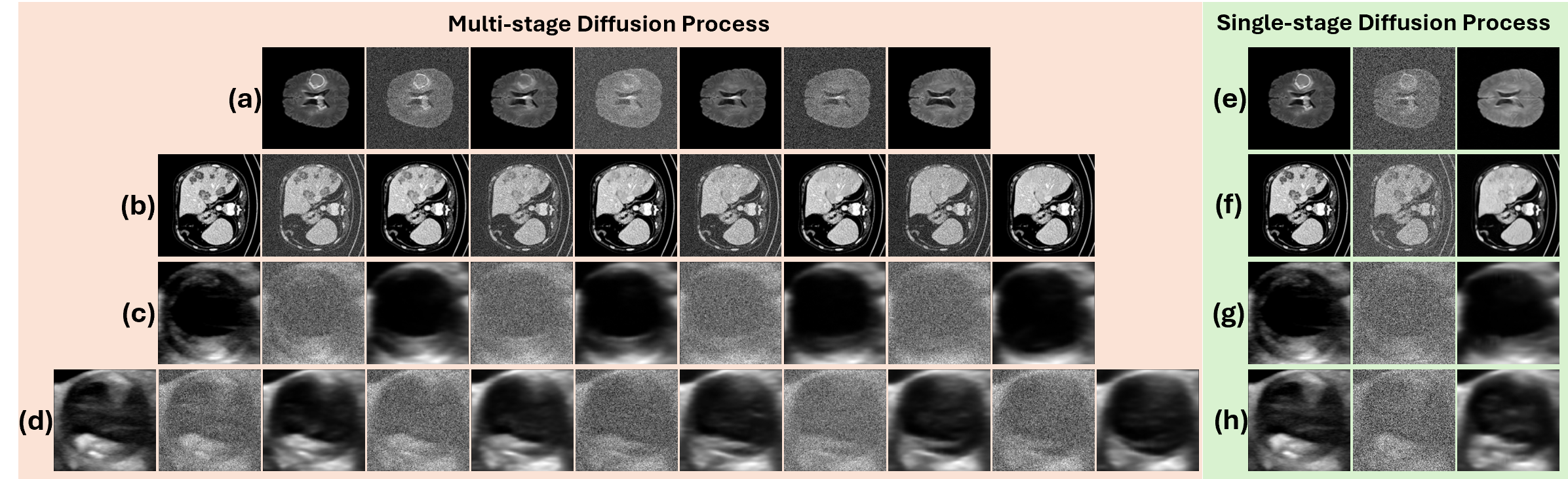}
    \caption{Visual comparison between images generated using multi-stage diffusion process and single-stage diffusion process with DDPM trained using Synomaly noise. The images are arranged in the order of input image, noised image, and denoised image. (a), (b), (c), and (d) depict images generated at each stage of the multi-stage diffusion process. (e), (f), (g), and (h) represent the corresponding images generated using the single-stage diffusion process.}
    \label{fig:ablation_multi}
\end{figure*}

\subsection{\revision{Image-level Anomaly Detection Performance}}
\revision{In this section, we evaluate the performance of the proposed method on image-level anomaly detection, assessing its ability to distinguish anomalous images from healthy ones. The evaluation is conducted using both healthy and anomalous images in the test set, and performance is measured using the Area Under the Receiver Operating Characteristic Curve (AUROC). The results are compared across different imaging modalities to demonstrate the method’s effectiveness in various clinical scenarios.}

\par
\revision{The AUROC scores achieved by the proposed method are 0.922 for brain MRI, 0.920 for liver CT, and 0.991 for carotid US, as illustrated in Fig.~\ref{fig:auroc_curves}. In the figure, the blue curve represents the AUROC for brain MRI, the orange curve corresponds to liver CT, and the green curve represents carotid US. These results indicate that the proposed method consistently performs well across different imaging modalities in distinguishing anomalous from healthy images. The particularly high AUROC score for carotid US (0.991) suggests that the method is highly effective in this modality, likely due to the more challenging appearance of anomalies compared to normal structures.}

\par
\revision{While these results confirm the strong classification capability of the proposed method, pixel-level anomaly segmentation provides additional benefits by offering localized information about abnormal regions, which can improve interpretability and aid in detailed analysis. Additionally, pixel-level AD is inherently more challenging than image-level classification, as it requires the model to distinguish fine-grained abnormal structures from normal variations within an image. Nonetheless, the inclusion of image-level classification metrics further supports the overall effectiveness of the approach and provides a broader evaluation perspective.}

\subsection{Ablation Study}
\revision{In this section, we conduct ablation studies to evaluate the impact of three key components of our method: the multi-stage diffusion process, Synomaly noise, and the hyperparameter selection for Synomaly noise. First, we analyze the effect of the multi-stage diffusion process by comparing it to the single-stage approach, assessing its role in improving anomaly removal while preserving fine details. Next, we evaluate the contribution of Synomaly noise to anomaly segmentation by comparing it with alternative noise functions. Finally, we examine the impact of hyperparameter selection for Synomaly noise, investigating how different configurations influence segmentation performance. These studies provide a deeper understanding of how each component contributes to the overall effectiveness of the proposed method.}

\subsubsection{Ablation for Multi-stage Diffusion Process}
To validate the effectiveness of the proposed multi-stage diffusion process, we conduct an ablation experiment to compare the anomaly segmentation performance with and without multi-stage diffusion. The DDPM was trained using Synomaly noise, and during inference, we tested the model's performance with a single-stage diffusion approach. To ensure a fair comparison, the optimal inference hyperparameters for single-stage diffusion with Synomaly noise were determined using a grid search and are reported in Tab.~\ref{tab_hyperparameter}, denoted as ``Synomaly+Single-stage''. In general, the number of noise steps at each stage for multi-stage inference is much smaller than the number of noise steps for single stage, which helps to preserve finer details.
\secRevision{The quantitative ablation results can be found in Tab.~\ref{tab_ablation_results_multi}, and the visual results can be found in Fig.~\ref{fig:ablation_multi}. It is notable that the reconstructed images shown in Fig.~\ref{fig:ablation_multi} are $\widehat{x_n}$, used for anomaly mask generation. Due to the generative nature of diffusion, they may exhibit gradual intensity changes even in undetected anomaly regions. However, in the multi-stage framework, the input to the next stage is $\widetilde{x_{n+1}}$, where background regions are restored from the original image to prevent over-modification.}

\begin{table}[ht!]
\caption{The ablation results on single-stage diffusion process and multi-stage diffusion process without masked fusion}\label{tab_ablation_results_multi}

\begin{threeparttable}
  \begin{tabular}{l ccc}
    \toprule
    \multirow{2}{*}{Metric} & Brain MRI & Liver CT & Carotid US \\
    \cline{2-4}
    & \multicolumn{3}{c}{\textbf{Synomaly+Single-stage}} \\
    DSC       & $0.738\pm0.226$ & $0.549\pm0.267$ & $0.688\pm0.126$ \\
    \secRevision{Precision} & \secRevision{$0.762\pm0.238$} & \secRevision{$0.584\pm0.281$} & \secRevision{$0.682\pm0.155$} \\
    Recall    & $0.747\pm0.243$ & $0.593\pm0.335$ & $0.694\pm0.168$ \\
    \hline
    & \multicolumn{3}{c}{\textbf{Synomaly+Multi-stage w/o Masked Fusion}} \\
    DSC       & $0.712\pm0.238$ & $0.539\pm0.277$ & $0.710\pm0.129$ \\
    \secRevision{Precision} & \secRevision{$0.774\pm0.267$} & \secRevision{$0.595\pm0.288$} & \secRevision{$0.733\pm0.169$} \\
    Recall    & $0.674\pm0.252$ & $0.551\pm0.313$ & $0.734\pm0.190$ \\
    \hline
    & \multicolumn{3}{c}{\textbf{Synomaly+Multi-stage w/ Masked Fusion}} \\
    DSC       & $0.739\pm0.234$ & $0.577\pm0.303$ & $0.731\pm0.117$ \\
    \secRevision{Precision} & \secRevision{$0.786\pm0.247$} & \secRevision{$0.622\pm0.302$} & \secRevision{$0.748\pm0.153$} \\
    Recall    & $0.741\pm0.259$ & $0.600\pm0.312$ & $0.772\pm0.180$ \\
    \bottomrule
  \end{tabular}
\end{threeparttable}
\end{table}

\begin{table*}[ht!]
\caption{\secRevision{Segmentation performance at each stage of the multi-stage diffusion process with masked fusion.}}\label{tab_results_multi_stage}
\resizebox{\textwidth}{!}{
\begin{threeparttable}
  \begin{tabular}{l ccc|ccc|ccc}
    \toprule
    &\multicolumn{3}{c}{\secRevision{Brain MRI BraTS23}}&\multicolumn{3}{c}{\secRevision{Liver CT LiTS}}&\multicolumn{3}{c}{\secRevision{Carotid Plaque US}}\\
    \cline{2-10}
     & \secRevision{DSC} & \secRevision{Precision} & \secRevision{Recall} & \secRevision{DSC} & \secRevision{Precision} & \secRevision{Recall} & \secRevision{DSC} & \secRevision{Precision} & \secRevision{Recall} \\
    \hline
    \secRevision{Stage 1} & \secRevision{$0.619\pm0.273$}          & \secRevision{$0.840\pm0.305$}          & \secRevision{$0.524\pm0.267$}          & \secRevision{$0.404\pm0.317$}          & \secRevision{$0.653\pm0.434$}         & \secRevision{$0.322\pm0.283$}          & \secRevision{$0.438\pm0.227$}          & \secRevision{$0.825\pm0.240$}          & \secRevision{$0.342\pm0.218$}\\
    \secRevision{Stage 2} & \secRevision{$0.710\pm0.251$}          & \secRevision{$0.831\pm0.260$}          & \secRevision{$0.660\pm0.270$}          & \secRevision{$0.507\pm0.319$}          & \secRevision{$0.697\pm0.369$}         & \secRevision{$0.446\pm0.320$}          & \secRevision{$0.651\pm0.160$}          & \secRevision{$0.832\pm0.155$}          & \secRevision{$0.586\pm0.209$}\\
    \secRevision{Stage 3} & \secRevision{$0.728\pm0.242$}          & \secRevision{$0.811\pm0.249$}          & \secRevision{$0.706\pm0.270$}          & \secRevision{$0.540\pm0.311$}          & \secRevision{$0.697\pm0.343$}         & \secRevision{$0.493\pm0.323$}          & \secRevision{$0.701\pm0.141$}          & \secRevision{$0.806\pm0.150$}          & \secRevision{$0.675\pm0.201$}\\
    \secRevision{Stage 4} & \secRevision{$0.737\pm0.231$}          & \secRevision{$0.800\pm0.238$}          & \secRevision{$0.731\pm0.261$}          & \secRevision{$0.551\pm0.293$}          & \secRevision{$0.665\pm0.311$}         & \secRevision{$0.527\pm0.318$}          & \secRevision{$0.715\pm0.132$}          & \secRevision{$0.779\pm0.151$}          & \secRevision{$0.716\pm0.197$}\\
    \secRevision{Stage 5} & \secRevision{$0.738\pm0.229$}          & \secRevision{$0.789\pm0.239$}          & \secRevision{$0.742\pm0.259$}          & \secRevision{$0.553\pm0.277$}          & \secRevision{$0.623\pm0.288$} 
            & \secRevision{$0.571\pm0.315$}          & \secRevision{$0.715\pm0.126$}          & \secRevision{$0.756\pm0.150$}          & \secRevision{$0.733\pm0.191$}\\
    \hline
  \end{tabular}
\end{threeparttable}
  }
\end{table*}

\par
For the case of brain MRI dataset, the single-stage DDPM trained using Synomaly noise successfully erases the round anomaly but struggles with generating a realistic lateral ventricle, as shown in Fig.~\ref{fig:ablation_multi}(e). The results using the multi-stage diffusion process show a more realistic reconstruction, particularly for the lateral ventricle, compared to the single-stage process. The detailed generation procedure of using multi-stage diffusion process can be found in Fig.~\ref{fig:multi_stage}(a). The anomalies are gradually removed, especially the abnormalities that appear in the area of lateral ventricle.
As depicted in Tab.~\ref{tab_ablation_results_multi}, the multi-stage diffusion process did not significantly improve anomaly segmentation in this case, showing similar performance compared to the single-stage approach. This is partly because the anomalies in FLAIR-weighted MRI are relatively clear, allowing the model to achieve satisfactory results with just a single iteration, making multiple diffusion iterations unnecessary. However, the comparison between Fig.~\ref{fig:ablation_multi}(a) and Fig.~\ref{fig:ablation_multi}(e) demonstrates that images generated using the multi-stage diffusion process are more realistic compared to those produced with a single diffusion step. 

\par
\begin{figure*}[ht!]
    \centering
    \includegraphics[width=0.88\textwidth]{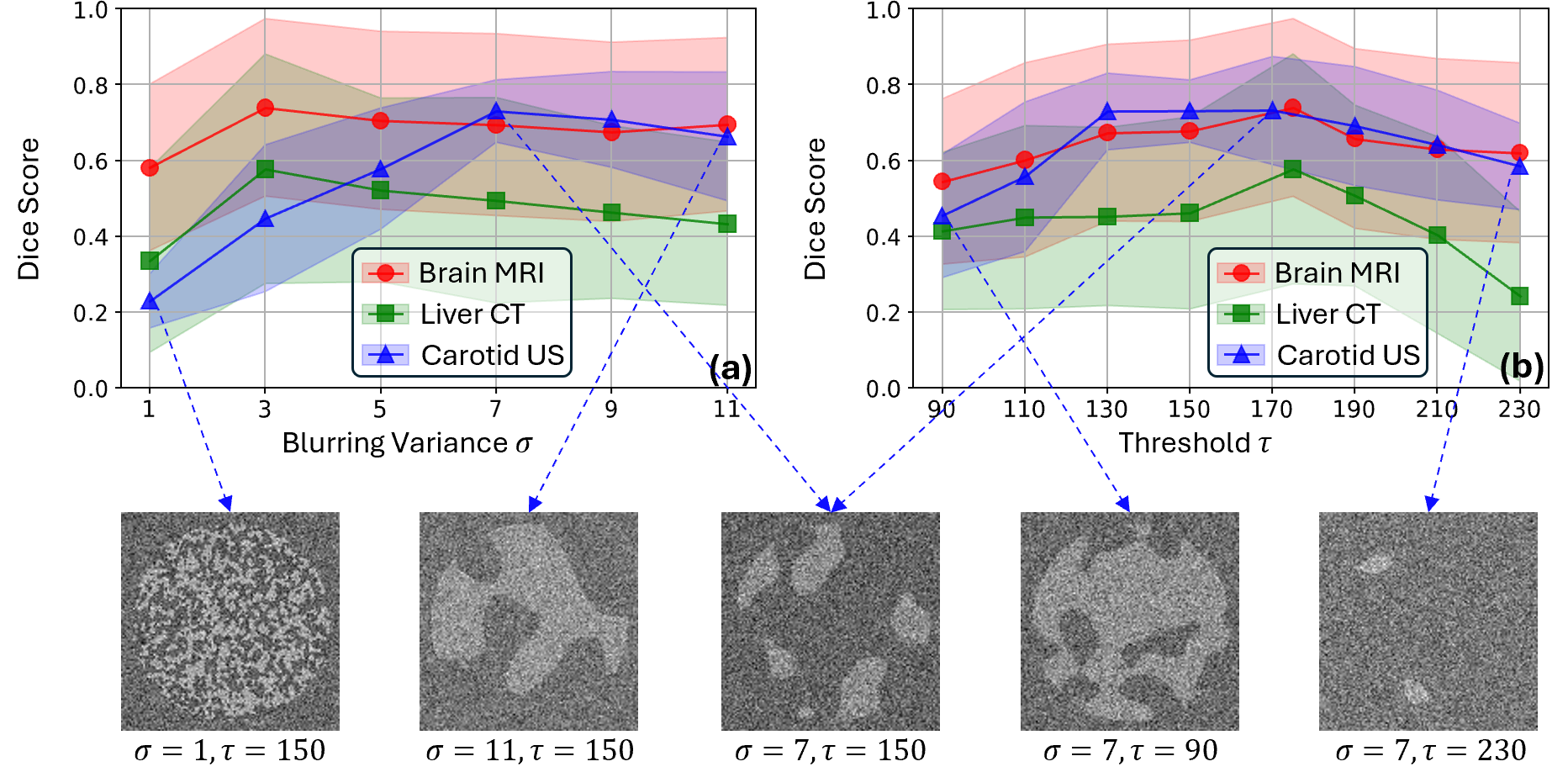}
    \caption{\revision{Ablation results of blurring variance ($\sigma$) and threshold value ($\tau$) for Synomaly noise on brain MRI, liver CT, and carotid plaque US data. The red, green, and blue lines represent the results for brain MRI, liver CT, and carotid US, respectively. The shaded areas indicate the standard deviation. Synomaly noise examples with different configurations are listed below.}}
    \label{fig:ablation_noise}
\end{figure*}

For the case of liver CT, the multi-stage diffusion process has successfully increased the Dice by around 0.03 compared to the single-stage approach. The visual results also support this observation. As depicted in Fig.~\ref{fig:ablation_multi}(f), the single-stage Synomaly method mostly removes the anomalies, but faint traces remain visible upon close inspection. In contrast, the multi-stage approach completely eliminates all the anomalies without a trace. Fig.~\ref{fig:ablation_noise}(b) shows how the abnormalities are progressively removed in each iteration of multi-stage diffusion process.

\par
For US carotid plaque anomaly segmentation, implementing the multi-stage diffusion process has improved the Dice score by approximately $0.05$ compared to the single-stage approach. The two visual examples in Fig.~\ref{fig:ablation_multi}(g) and Fig.~\ref{fig:ablation_multi}(h) also demonstrate that a single diffusion process cannot fully erase the anomalous regions. While the plaque in the upper part of the vessel is successfully removed, the larger plaque at the bottom of the vessel remains in the generated image. After applying multi-stage diffusion process, all the plaques are fully removed. Fig.~\ref{fig:ablation_multi}(c) shows the generated images at each stage, clearly illustrating how the plaques are iteratively removed from the original image. Generally, distinct and large anomalies require more iterations to be completely erased. Fig.~\ref{fig:ablation_multi}(d) showcases another example involving five iterations, where the plaque at the bottom of the vessel lumen is progressively removed over five iterations, while the smaller plaque at the top is mostly removed in four iterations.

\par
\secRevision{The benefit of the multi-stage diffusion process is most evident in carotid US, where anomalies are more distinct and require stronger corruption to be removed. The multi-stage approach, combined with masked fusion, helps suppress these anomalies while preserving healthy regions. In contrast, for brain MRI and liver CT, where small noise levels are already sufficient to blur the anomaly, the improvement is less significant.}

\par
\secRevision{To further evaluate the effectiveness of the proposed multi-stage diffusion process, we calculate the average DSC, precision, and recall of the intermediate anomaly masks at each stage, as summarized in Tab.~\ref{tab_results_multi_stage}. For this analysis, we include only those test images that underwent five diffusion stages during inference. As shown in Tab.~\ref{tab_results_multi_stage}, across the multi-stage diffusion process, recall consistently increases, while precision gradually decreases, and DSC improves and eventually stabilizes. This indicates that the model progressively captures more of the anomaly regions over successive stages, resulting in higher recall. The decrease in precision reflects a broader predicted mask that includes more false positives, but the gain in recall outweighs this, leading to improved DSC. By stage four and five, all metrics begin to converge, suggesting that the refinement process reaches a stable segmentation output.}

\par
Additionally, we evaluated the performance without using the masked fusion technique, where the generated image from the previous iteration was directly used as input for the next iteration of the diffusion model. The results, labeled as ``Synomaly+Multi-stage w/o Masked Fusion'', are shown in Tab.~\ref{tab_ablation_results_multi}. Compared to the results of applying masked fusion technique, labeled as ``Synomaly+Multi-stage w/ Masked Fusion'', the masked fusion led to an increase of approximately 0.02 in Dice score across all datasets. Without masked fusion, the non-anomalous parts of the image also undergo multiple diffusion processes, which may introduce changes to healthy regions. By using the masked fusion technique, the original texture in non-diseased areas is preserved, thereby enhancing anomaly segmentation accuracy.

\par
\revision{Lastly, we calculated the computational time for both single- and multi-stage diffusion processes. The proposed multi-stage diffusion process introduces additional computational steps compared to single-stage diffusion, resulting in an inference time of 67.6$\pm$5.7 milliseconds per image, compared to 13.8$\pm$5.6 milliseconds for the single-stage approach. The inference speed for other diffusion-based methods with single-stage diffusion is similar to our single-stage implementation. Notably, 67.6 milliseconds per image corresponds to approximately 15 frames per second, which is sufficient for most real-time applications. This ensures that the method remains computationally feasible for practical use while enabling effective anomaly detection.}

\subsubsection{Ablation for Synomaly Noise}
\revision{To evaluate the effectiveness of Synomaly noise, we compare the performance of different noise functions when only a single-stage diffusion process is applied. The noise functions considered include Gaussian, Coarse, Pyramid, and Simplex noise, with the corresponding results presented in Tab.~\ref{tab_results} as DDPM+Gaussian, DDPM+Coarse, DDPM+Pyramid, and DDPM+Simplex, respectively. The performance of Synomaly noise in a single-stage diffusion process is reported in Tab.~\ref{tab_ablation_results_multi} as Synomaly+Single-stage.}

\par
\revision{For brain tumor segmentation, Synomaly noise achieved a Dice score of 0.738, outperforming other noise functions by at least 0.1. Similarly, for carotid plaque segmentation, Synomaly noise yielded a Dice score of 0.688, surpassing the best alternative by over 0.3. In liver tumor segmentation, Synomaly noise alone outperformed the best-performing SOTA noise function (Simplex noise) by 0.01 in Dice score, demonstrating its superiority across all datasets. While the performance gain in liver CT was less pronounced compared to the other datasets, Synomaly noise still consistently outperformed existing noise functions. These results highlight the advantages of Synomaly noise in generating realistic counterfactual images and enhancing anomaly segmentation performance.}

\subsubsection{Hyperparameter Ablation for Synomaly Noise}
In this ablation study, we examine the influence of the hyperparameters of Synomaly noise, as described in Sec.~\ref{sec:synomaly}. The performance of Synomaly noise is closely linked to its ability to represent the target anomalies. If the Synomaly noise generated during training resembles the real abnormalities encountered during inference, the trained model can more effectively erase the diseased areas during testing. The morphology of the generated Synomaly noise is controlled by two parameters: the blurring variance ($\sigma$) and the threshold value ($\tau$).

\par
\revision{To evaluate the influence of these parameters, several models were trained with different values of $\sigma$ and $\tau$ on the brain MRI, liver CT, and carotid US dataset, respectively.} The generated noise is normalized to the range [0, 255], and consequently, the threshold values are also within this range. 
\revision{For brain MRI and liver CT, the threshold value ($\tau$) was fixed at 175, while the blurring variance ($\sigma$) varied from 1 to 11. For carotid US, $\tau$ was fixed at 150, with $\sigma$ also ranging from 1 to 11. The results are shown in Fig.~\ref{fig:ablation_noise}(a), where the red, green, and blue lines represent brain MRI, liver CT, and carotid US, respectively.
It is evident that the optimal performance is achieved at $\sigma=3$ for brain MRI and liver CT, whereas for carotid US, the best choice is $\sigma=7$.
Similarly, for $\tau$, we conducted experiments by fixing $\sigma$ at 3 while varying $\tau$ from 90 to 230 for MRI and CT data. For US, $\sigma$ was fixed at 7. The results, as shown in Fig.~\ref{fig:ablation_noise}(b), indicate that the best performance is achieved when setting $\tau$ to 175 for MRI and CT, while $\tau=150$ is the best choice for carotid US.}
Some examples of the generated Synomaly noise are displayed below the plots. The value of $\sigma$ controls the size of the synthetic anomaly regions, while $\tau$ determines the number of generated artificial anomaly areas. A smaller $\sigma$ value results in smaller synthetic anomaly regions, whereas a lower $\tau$ value leads to a higher number of generated anomaly areas. However, in some cases, when choosing a smaller $\tau$ value, these generated regions sometimes overlap with each other, leading to fewer but larger anomalous areas.
\revision{According to this ablation study, the optimal parameters for brain MRI and liver CT data are $\sigma = 3$, $\tau = 175$, while the best combination for carotid US is $\sigma = 7$, $\tau = 150$, since these parameter choices can best resemble the morphologies of corresponding anomalies based on our experiments.}

\revision{To facilitate hyperparameter selection without an exhaustive grid search, we provide practical guidelines based on the relative size of anomalous regions. For a 128$\times$128 image, we recommend $\sigma=1,~\tau=180$ for small anomalies ($\sim < 2\%$ of the image), while for moderate anomalies ($\sim 2\%-5\%$), such as brain and liver tumor a setting of $\sigma=3,~\tau=175$ is suggested. Intermediate cases ($\sim 5\%-7.5\%$) can use $\sigma=5,~\tau=160$ as a reasonable starting point. For larger anomalies ($\sim 10\%$), such as carotid US plaques, $\sigma=7,~\tau=150$ is a suitable choice. These anchor points serve as a structured reference for hyperparameter selection. A smaller $\sigma$ generates smaller synthetic anomaly regions, while a lower $\tau$ increases the number of generated anomaly areas. By following these guidelines, users can refine parameters accordingly, reducing computational overhead while maintaining robust anomaly detection performance. 
Due to the random size distribution of the generated synthetic anomalies (see Fig.~\ref{fig:ablation_noise} lower half), a model trained with a single parameter setting is not strictly limited to detecting anomalies of a specific size but can still generalize to certain out-of-distribution cases.}

\section{\secRevision{Conclusions}}
In this paper, we presented a novel framework for unsupervised anomaly detection in \revision{medical images} using Synomaly noise and a multi-stage diffusion process. Synomaly noise was designed to add synthetic anomalies to healthy images during training, enabling the model to effectively remove anomalies during inference. The multi-stage diffusion process was introduced to incrementally denoise the images, thereby preserving fine details and generating more realistic reconstructions.

\par
The experiments on carotid US, brain MRI, and liver CT datasets demonstrated the efficacy of the proposed approach. For the carotid US dataset, the proposed method achieved performance comparable to fully supervised segmentation models, demonstrating its strong anomaly removal capabilities even without large amounts of labeled data. The proposed method improved unsupervised anomaly segmentation results by 0.23 in the carotid US dataset. The brain MRI experiments showed an increase of approximately 0.1 in Dice score with the proposed approach, while the liver CT dataset demonstrated an increase of 0.04, showing improved unsupervised anomaly segmentation accuracy. Additionally, the ablation studies highlighted the importance of configuring reasonable hyperparameters for Synomaly noise, as its performance is closely tied to its ability to resemble the target anomalies. \revision{Our experiments also showed that Synomaly noise alone improves anomaly segmentation performance compared to other noise functions, indicating its effectiveness in generating counterfactual images. We also evaluated the multi-stage diffusion process and masked fusion technique, which contributed to further improvements in segmentation quality. These findings suggest that both Synomaly noise and multi-stage diffusion play important roles in the overall performance of the proposed method.}

\par
Overall, the proposed approach outperformed existing state-of-the-art methods for unsupervised anomaly detection, significantly narrowing the gap between unsupervised and fully supervised approaches. Despite challenges such as variability in acquisition settings and machine types, the proposed method exhibited robustness, generalizability, and promising potential for clinical applications. Moreover, the proposed framework provides an opportunity for doctors to incorporate their prior knowledge of the disease's appearance, thus enhancing unsupervised anomaly segmentation accuracy by allowing the model to be guided by expert input. \revision{In the future, hyperparameter selection can also be automated for non-expert users by incorporating a few segmentation examples.}
Further research could also focus on could also explore training the model to selectively detect specific types of anomalies while disregarding others, enabling selective segmentation. Such advancements would allow the model not only to detect anomalies but also to classify them to some extent, thereby adding greater value to clinical decision-making and facilitating more tailored medical interventions.

\par
\secRevision{The proposed approach is particularly well-suited for addressing anomalies that are distinguishable based on intensity differences. Anomalies that present clear intensity variations compared to surrounding tissues are efficiently handled by this framework, resulting in high segmentation accuracy. However, the approach is less effective when dealing with anomalies that do not have distinctive intensity contrasts but instead differ in shape or structural characteristics. In such cases, the proposed model does not show evident superiority to differentiate between normal and abnormal shapes, as the generative process is primarily guided by intensity-based information.} \final{Recent studies have explored integrating structural priors~\citep{bercea2025evaluating}, synthetic perturbations~\citep{marimont2024ensembled}, and language-guided representations~\citep{zhou2025ultraad} to enhance robustness in such scenarios. Future work could explore combining intensity-based anomaly detection with additional spatial or shape features to extend the model’s applicability to a broader range of complex anomalies~\citep{bercea2025nova}.}

\section*{Declaration of Competing Interest}
\label{competinginterest}

The authors report no conflicts of interest.

\section*{Statement}
During the preparation of this work the author(s) used ChatGPT in order to revise the manuscript for fluent and grammatically correct English. After using this tool/service, the author(s) reviewed and edited the content as needed and take(s) full responsibility for the content of the published article.


\setcounter{section}{1}  
\renewcommand{\thesection}{\Alph{section}}        
\renewcommand{\thesubsection}{\thesection.\arabic{subsection}}  
\renewcommand{\thefigure}{A1}

\section*{\secRevision{Appendix A. Dataset Details}}
\begin{figure}[ht!]
    \centering
    \includegraphics[width=0.48\textwidth]{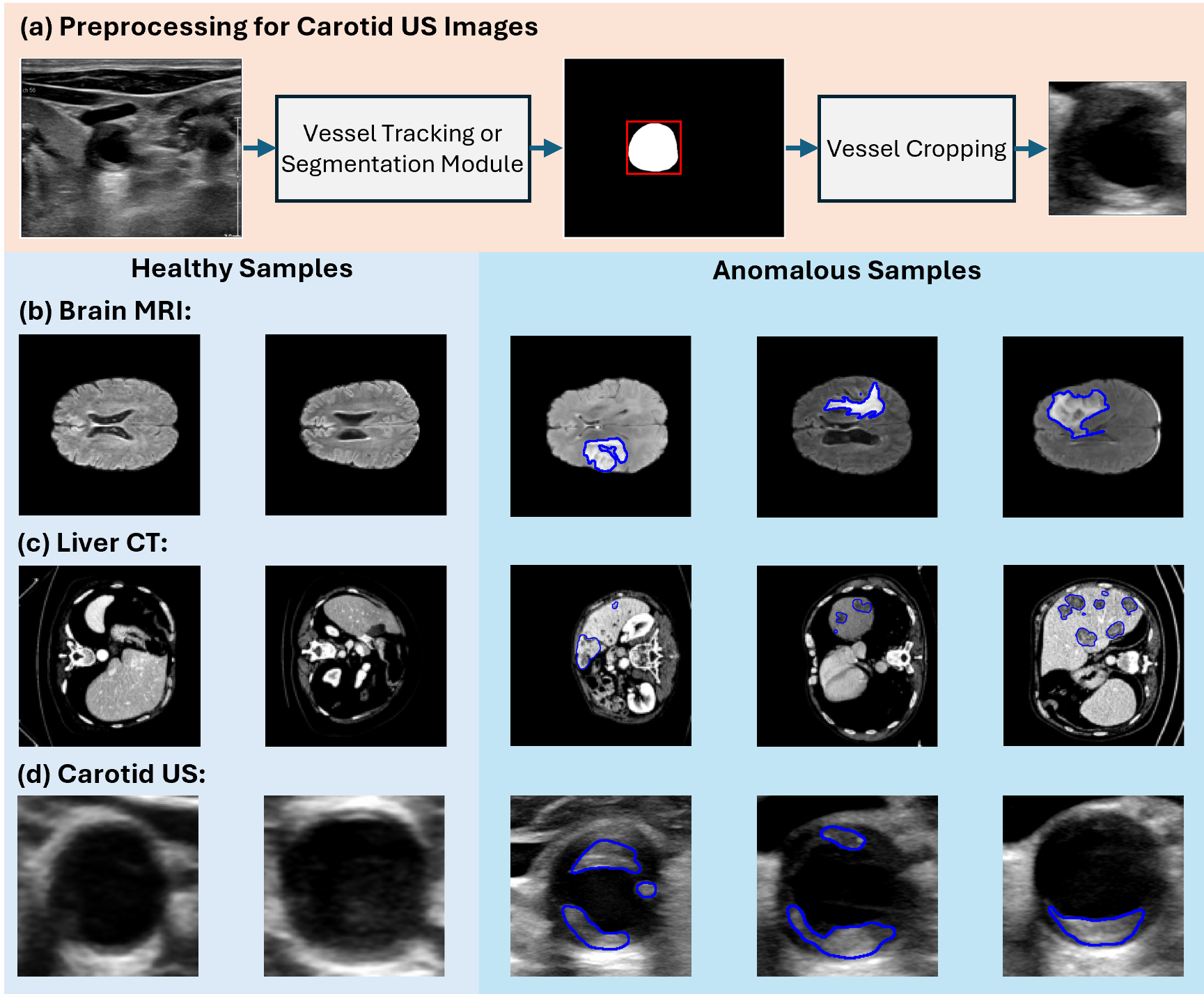}
    \caption{\secRevision{Dataset overview. The anomalous regions are marked using blue boundaries.}}
    \label{fig:dataset}
\end{figure}

\subsection{\secRevision{Brain MRI: BraTS Dataset}}
\secRevision{The BraTS2023 dataset, from the international Brain Tumor Segmentation Challenge~\citep{baid2021rsna, bakas2017advancing, menze2014multimodal}, is a widely used benchmark in medical imaging. It includes 3D brain MRI scans from $1251$ tumor patients with pixel-wise annotations for healthy tissue, GD-enhancing tumor (ET), edematous/invaded tissue (ED), and necrotic tumor core (NCR). In this work, all tumor-related labels (ET, ED, NCR) are combined into a single ``anomalous'' class, yielding two categories: healthy and anomalous tissue.}

\par
\secRevision{Since the focus is on tumor detection in 2D images, only axial brain slices are considered in this work and the 2D tumor ground truth segmentation is extracted from the annotated 3D classification volumes. Given that ET, ED and NCR are most commonly found in the cerebral hemispheres, the brain MRI dataset is further refined to include only the central slices. For anomaly detection, the FLAIR sequence is used. All selected slices are resized to $128\times 128$ pixels to have unified dimensions. $1000$ anomalous slices are randomly selected as the test set.
In total, the dataset consists of $3743$ healthy training samples, $1000$ anomalous test samples, and $416$ healthy test samples. \secRevision{For normalization, we first clip the intensity values at the 99th percentile to mitigate the effect of extreme outliers, and then apply min-max normalization.}
To help intuitive understanding, a few representative samples are visualized in Fig.~\ref{fig:dataset}(b).}

\subsection{\secRevision{Liver CT: LiTS Dataset}}
\secRevision{The Liver Tumor Segmentation Benchmark (LiTS) dataset \citep{bilic2023liver} includes $130$ abdominal CT volumes, each with annotations of the liver and tumors. For the purposes of this work, all tumors are categorized as ``anomalous'' class.}

\par
\secRevision{Each 3D abdominal CT volume contains varying axial slices, ranging from $42$ and $1026$. To prepare the data for model training and evaluation, the annotated segmentations are used to classify slices into healthy and anomalous cases. Further refinement is performed using size thresholding to isolate the relevant slices and all images are resized to $128\times128$ pixels. 
In total, the CT dataset has $5820$ healthy training images, $1000$ anomalous test images, and $647$ healthy test images. \secRevision{To normalize the images, we first clip the intensity values at the 99th percentile to reduce the impact of extreme outliers, followed by min-max scaling to the [0,1] range.} 
A set of representative samples is shown in Fig.~\ref{fig:dataset}(c). These examples highlight the complexity of the anomaly detection task, as the abdominal region exhibits various appearances and the liver itself can differ in size, shape and location. The relatively small ratio between the anomaly and the chest volume can further cause class imbalance problems. 
These factors make it particularly challenging to accurately segment liver tumors, while preserving healthy tissue. This high diversity and complexity of the LiTS dataset make it a great benchmark for assessing the performance and generalizability of the proposed anomaly detection approaches.}

\subsection{\secRevision{Carotid US Dataset}}
\secRevision{The carotid artery US dataset includes US scans from $23$ healthy patients and $13$ patients with carotid plaques.
The US images in the dataset are all captured from the cross-sectional view of the carotid artery. The acquisition was performed within the Institutional Review Board Approval by the Ethical Commission of the Technical University of Munich (reference number 2025-60-S-CB).
Due to the nature of US imaging, B-mode appearance is affected significantly by contact point, force, angle, and US machine setting. So, to alleviate the effects caused by the changing image appearance, we crop the US image around the target anatomy, i.e., the carotid artery, to ensure the network only focuses on disease-cased image variations. 
In this study, we applied existing vessel tracking and segmentation methods to generate vessel masks, which are then used to crop the original full-view US images. Fig.~\ref{fig:dataset}(a) presents the pipeline for vessel tracking and cropping procedure. To account for potential imperfections in vessel segmentation and to enhance model robustness, a small degree of randomness regarding the position and size of the vessel within the image is introduced in the cropping process, which also creates some variability in the training data.}

\par
\secRevision{The ground truth annotations for plaques in the anomalous carotid artery samples are manually annotated under the supervision of our clinical experts specialized in vascular disease. All such annotations of vessels and anomalies were performed using the ImFusion Suite\footnote{\url{https://www.imfusion.com/products/imfusion-suite}}.
All US images are resized to $128\times128$ pixels to standardize input dimensions. \secRevision{A min-max normalization is applied to normalize the images to [0,1].}
In total, the final dataset contains $7306$ healthy training images, $545$ anomalous test images, and $833$ healthy test images.
The examples can be found in Fig.~\ref{fig:dataset}(d).}

\begin{strip}

\section*{\secRevision{Appendix B. Additional Visual Results}}

\raggedright
\setlength{\parindent}{1em}
\secRevision{Fig.~\ref{fig:experiments_results_brats}, \ref{fig:experiments_results_lits}, and \ref{fig:experiments_results_us} present additional visual results on some challenging cases in the three datasets.}

\centering
\includegraphics[width=0.95\textwidth]{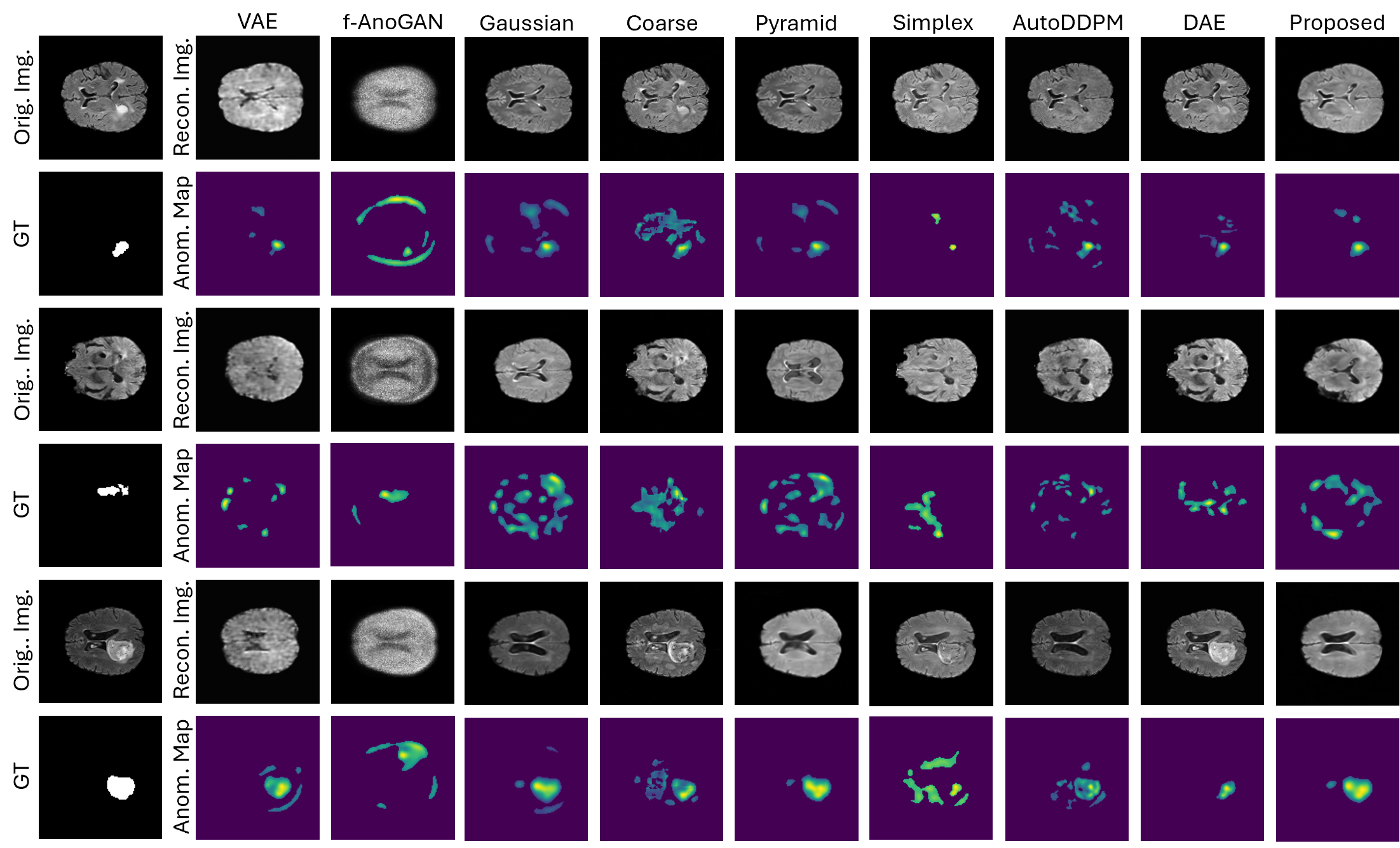}
\renewcommand{\thefigure}{B1}
\captionof{figure}{\secRevision{Extra challenging samples of anomaly segmentation results using different unsupervised methods on brain MRI dataset.}}
\label{fig:experiments_results_brats}

\justifying
\setlength{\parindent}{1em}
\secRevision{Fig.~\ref{fig:experiments_results_brats} shows three extra visual results from the brain MRI dataset. The first two examples illustrate failure cases of the proposed method. In the first, the model incorrectly identifies a small bright region in the upper part of the image as an anomaly. Increasing the $\sigma$ value and decreasing the $\tau$ value may improve robustness by making the model less sensitive to small artifacts and more focused on larger anomalous regions. In the second case, the failure is primarily due to the out-of-distribution appearance of the input, which deviates significantly from the training distribution. Incorporating a broader variety of training samples may help mitigate such issues. Nevertheless, the proposed method still demonstrates superior performance compared to other SOTA approaches, even in these challenging cases.}

\centering
\includegraphics[width=0.94\textwidth]{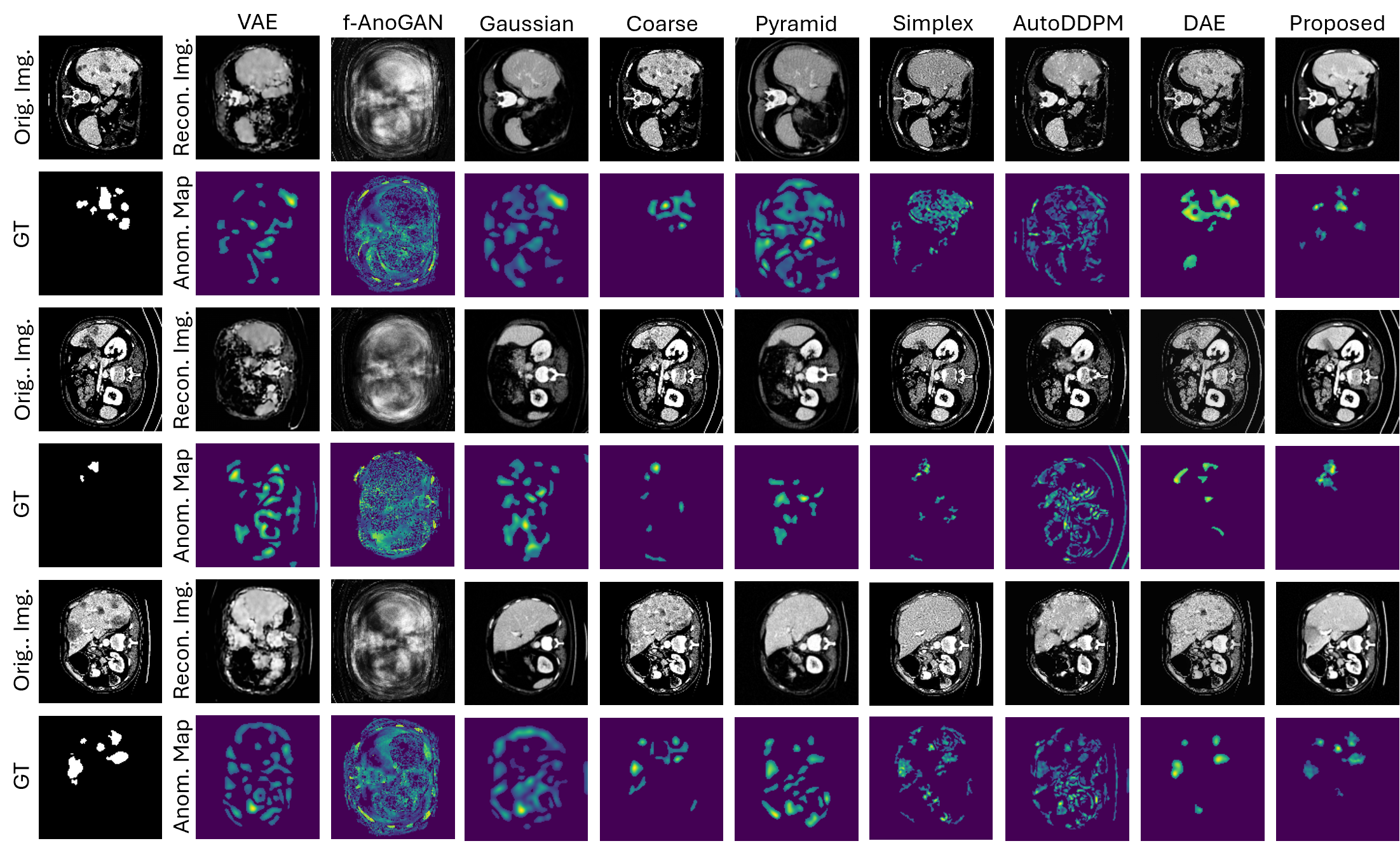}
\renewcommand{\thefigure}{B2}
\captionof{figure}{\secRevision{Extra challenging samples of anomaly segmentation results using different unsupervised methods on liver CT dataset.}}
\label{fig:experiments_results_lits}

\justifying
\setlength{\parindent}{1em}
\secRevision{Fig.~\ref{fig:experiments_results_lits} presents three additional examples from the liver CT dataset. In the first example, the model fails to remove three lesions, likely mistaking them for small hepatic ducts, resulting in missed anomaly detection. Decreasing $\sigma$ and increasing $\tau$ during training could improve sensitivity to such subtle lesions. In the second case, the gastrointestinal duct appears as an irregular structure in the lower left but is correctly reconstructed and not mistaken as an anomaly. However, a dark region in the liver is wrongly identified, causing false positives. The third example shows good localization, with most anomalies accurately detected. Despite these challenges, the proposed method consistently outperforms other SOTA approaches.}

\centering
\includegraphics[width=0.94\textwidth]{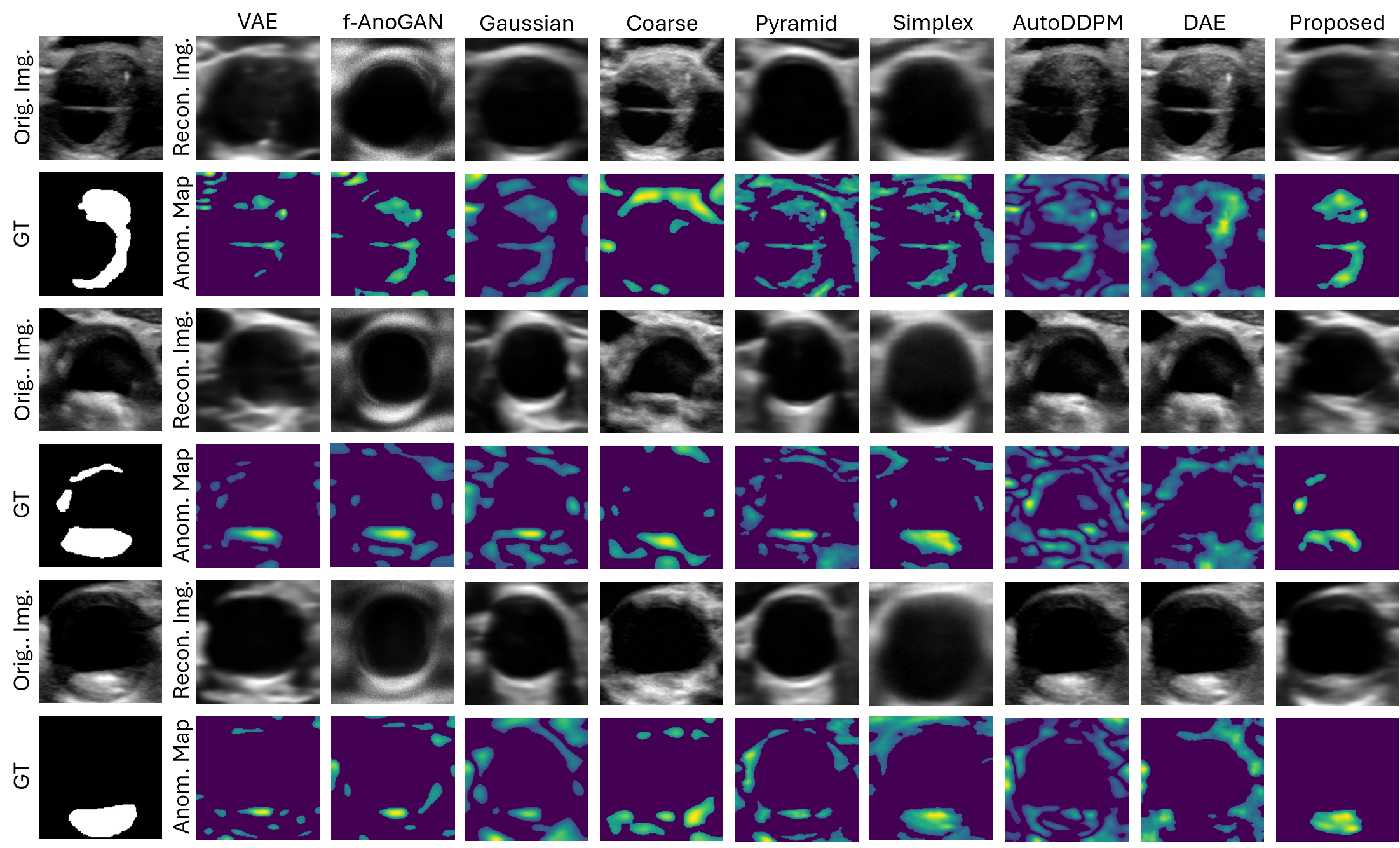}
\renewcommand{\thefigure}{B3}
\captionof{figure}{\secRevision{Extra challenging samples of anomaly segmentation results using different unsupervised methods on carotid US dataset.}}
\label{fig:experiments_results_us}

\justifying
\setlength{\parindent}{1em}
\secRevision{Fig.~\ref{fig:experiments_results_us} presents three additional examples from the carotid US dataset. In the first image, an artifact within the carotid lumen is incorrectly identified as an anomaly and removed in the generated counterfactual image, resulting in a false positive prediction. In the second case, the plaque exhibits a highly distinct boundary and appearance. While the proposed method successfully detects some smaller plaques in the upper region, the larger and more prominent plaque at the bottom is not fully removed. A similar issue is observed in the third example, where the distinct appearance of the plaque hinders its complete removal. To improve the model’s performance on such distinct anomalies, increasing the intensity offset $i$ during training can produce more distinct Synomaly noise, enhancing the model’s ability to detect and remove pronounced anomalies. Despite these challenges, the proposed method still demonstrates robust performance compared to other SOTA approaches.}
\end{strip}

\bibliographystyle{model2-names.bst}\biboptions{authoryear}
\bibliography{references}

\begin{thebibliography}{64}
\expandafter\ifx\csname natexlab\endcsname\relax\def\natexlab#1{#1}\fi
\providecommand{\url}[1]{\texttt{#1}}
\providecommand{\href}[2]{#2}
\providecommand{\path}[1]{#1}
\providecommand{\DOIprefix}{doi:}
\providecommand{\ArXivprefix}{arXiv:}
\providecommand{\URLprefix}{URL: }
\providecommand{\Pubmedprefix}{pmid:}
\providecommand{\doi}[1]{\href{http://dx.doi.org/#1}{\path{#1}}}
\providecommand{\Pubmed}[1]{\href{pmid:#1}{\path{#1}}}
\providecommand{\bibinfo}[2]{#2}
\ifx\xfnm\relax \def\xfnm[#1]{\unskip,\space#1}\fi
\bibitem[{Akcay et~al.(2018)Akcay, Atapour-Abarghouei and Breckon}]{akcay2018ganomaly}
\bibinfo{author}{Akcay, S.}, \bibinfo{author}{Atapour-Abarghouei, A.}, \bibinfo{author}{Breckon, T.P.}, \bibinfo{year}{2018}.
\newblock \bibinfo{title}{Ganomaly: Semi-supervised anomaly detection via adversarial training}, in: \bibinfo{booktitle}{Asian conference on computer vision}, \bibinfo{organization}{Springer}. pp. \bibinfo{pages}{622--637}.
\bibitem[{Baid et~al.(2021)Baid, Ghodasara, Mohan, Bilello, Calabrese, Colak, Farahani, Kalpathy-Cramer, Kitamura, Pati et~al.}]{baid2021rsna}
\bibinfo{author}{Baid, U.}, \bibinfo{author}{Ghodasara, S.}, \bibinfo{author}{Mohan, S.}, \bibinfo{author}{Bilello, M.}, \bibinfo{author}{Calabrese, E.}, \bibinfo{author}{Colak, E.}, \bibinfo{author}{Farahani, K.}, \bibinfo{author}{Kalpathy-Cramer, J.}, \bibinfo{author}{Kitamura, F.C.}, \bibinfo{author}{Pati, S.}, et~al., \bibinfo{year}{2021}.
\newblock \bibinfo{title}{The rsna-asnr-miccai brats 2021 benchmark on brain tumor segmentation and radiogenomic classification}.
\newblock \bibinfo{journal}{arXiv preprint arXiv:2107.02314} .
\bibitem[{Bakas et~al.(2017)Bakas, Akbari, Sotiras, Bilello, Rozycki, Kirby, Freymann, Farahani and Davatzikos}]{bakas2017advancing}
\bibinfo{author}{Bakas, S.}, \bibinfo{author}{Akbari, H.}, \bibinfo{author}{Sotiras, A.}, \bibinfo{author}{Bilello, M.}, \bibinfo{author}{Rozycki, M.}, \bibinfo{author}{Kirby, J.S.}, \bibinfo{author}{Freymann, J.B.}, \bibinfo{author}{Farahani, K.}, \bibinfo{author}{Davatzikos, C.}, \bibinfo{year}{2017}.
\newblock \bibinfo{title}{Advancing the cancer genome atlas glioma mri collections with expert segmentation labels and radiomic features}.
\newblock \bibinfo{journal}{Scientific data} \bibinfo{volume}{4}, \bibinfo{pages}{1--13}.
\bibitem[{Baur et~al.(2021)Baur, Denner, Wiestler, Navab and Albarqouni}]{baur2021autoencoders}
\bibinfo{author}{Baur, C.}, \bibinfo{author}{Denner, S.}, \bibinfo{author}{Wiestler, B.}, \bibinfo{author}{Navab, N.}, \bibinfo{author}{Albarqouni, S.}, \bibinfo{year}{2021}.
\newblock \bibinfo{title}{Autoencoders for unsupervised anomaly segmentation in brain mr images: a comparative study}.
\newblock \bibinfo{journal}{Medical Image Analysis} \bibinfo{volume}{69}, \bibinfo{pages}{101952}.
\bibitem[{Baur et~al.(2019)Baur, Wiestler, Albarqouni and Navab}]{baur2019deep}
\bibinfo{author}{Baur, C.}, \bibinfo{author}{Wiestler, B.}, \bibinfo{author}{Albarqouni, S.}, \bibinfo{author}{Navab, N.}, \bibinfo{year}{2019}.
\newblock \bibinfo{title}{Deep autoencoding models for unsupervised anomaly segmentation in brain mr images}, in: \bibinfo{booktitle}{Brainlesion: Glioma, Multiple Sclerosis, Stroke and Traumatic Brain Injuries: 4th International Workshop, BrainLes 2018, Held in Conjunction with MICCAI 2018, Granada, Spain, September 16, 2018, Revised Selected Papers, Part I 4}, \bibinfo{organization}{Springer}. pp. \bibinfo{pages}{161--169}.
\bibitem[{Behrendt et~al.(2024)Behrendt, Bhattacharya, Kr{\"u}ger, Opfer and Schlaefer}]{behrendt2024patched}
\bibinfo{author}{Behrendt, F.}, \bibinfo{author}{Bhattacharya, D.}, \bibinfo{author}{Kr{\"u}ger, J.}, \bibinfo{author}{Opfer, R.}, \bibinfo{author}{Schlaefer, A.}, \bibinfo{year}{2024}.
\newblock \bibinfo{title}{Patched diffusion models for unsupervised anomaly detection in brain mri}, in: \bibinfo{booktitle}{Medical Imaging with Deep Learning}, \bibinfo{organization}{PMLR}. pp. \bibinfo{pages}{1019--1032}.
\bibitem[{Bercea et~al.(2025a)Bercea, Li, Raffler, Riedel, Schmitzer, Kurz, Bitzer, Ro{\ss}m{\"u}ller, Canisius, Beyrle et~al.}]{bercea2025nova}
\bibinfo{author}{Bercea, C.I.}, \bibinfo{author}{Li, J.}, \bibinfo{author}{Raffler, P.}, \bibinfo{author}{Riedel, E.O.}, \bibinfo{author}{Schmitzer, L.}, \bibinfo{author}{Kurz, A.}, \bibinfo{author}{Bitzer, F.}, \bibinfo{author}{Ro{\ss}m{\"u}ller, P.}, \bibinfo{author}{Canisius, J.}, \bibinfo{author}{Beyrle, M.L.}, et~al., \bibinfo{year}{2025}a.
\newblock \bibinfo{title}{Nova: A benchmark for anomaly localization and clinical reasoning in brain mri}.
\newblock \bibinfo{journal}{arXiv preprint arXiv:2505.14064} .
\bibitem[{Bercea et~al.(2023)Bercea, Neumayr, Rueckert and Schnabel}]{bercea2023mask}
\bibinfo{author}{Bercea, C.I.}, \bibinfo{author}{Neumayr, M.}, \bibinfo{author}{Rueckert, D.}, \bibinfo{author}{Schnabel, J.A.}, \bibinfo{year}{2023}.
\newblock \bibinfo{title}{Mask, stitch, and re-sample: Enhancing robustness and generalizability in anomaly detection through automatic diffusion models}, in: \bibinfo{booktitle}{ICML 3rd Workshop on Interpretable Machine Learning in Healthcare (IMLH)}.
\bibitem[{Bercea et~al.(2024)Bercea, Wiestler, Rueckert and Schnabel}]{bercea2024diffusion}
\bibinfo{author}{Bercea, C.I.}, \bibinfo{author}{Wiestler, B.}, \bibinfo{author}{Rueckert, D.}, \bibinfo{author}{Schnabel, J.A.}, \bibinfo{year}{2024}.
\newblock \bibinfo{title}{Diffusion models with implicit guidance for medical anomaly detection}, in: \bibinfo{booktitle}{Proceedings of the 27th International Conference on Medical Image Computing and Computer-Assisted Intervention (MICCAI 2024)}, \bibinfo{publisher}{Springer Nature Switzerland}. pp. \bibinfo{pages}{211--220}.
\bibitem[{Bercea et~al.(2025b)Bercea, Wiestler, Rueckert and Schnabel}]{bercea2025evaluating}
\bibinfo{author}{Bercea, C.I.}, \bibinfo{author}{Wiestler, B.}, \bibinfo{author}{Rueckert, D.}, \bibinfo{author}{Schnabel, J.A.}, \bibinfo{year}{2025}b.
\newblock \bibinfo{title}{Evaluating normative representation learning in generative ai for robust anomaly detection in brain imaging}.
\newblock \bibinfo{journal}{Nature Communications} \bibinfo{volume}{16}, \bibinfo{pages}{1624}.
\bibitem[{Bi et~al.(2023)Bi, Jiang, Clarenbach, Ghotbi, Karlas and Navab}]{bi2023mi}
\bibinfo{author}{Bi, Y.}, \bibinfo{author}{Jiang, Z.}, \bibinfo{author}{Clarenbach, R.}, \bibinfo{author}{Ghotbi, R.}, \bibinfo{author}{Karlas, A.}, \bibinfo{author}{Navab, N.}, \bibinfo{year}{2023}.
\newblock \bibinfo{title}{Mi-segnet: Mutual information-based us segmentation for unseen domain generalization}, in: \bibinfo{booktitle}{International Conference on Medical Image Computing and Computer-Assisted Intervention}, \bibinfo{organization}{Springer}. pp. \bibinfo{pages}{130--140}.
\bibitem[{Bi et~al.(2024)Bi, Jiang, Duelmer, Huang and Navab}]{bi2024machine}
\bibinfo{author}{Bi, Y.}, \bibinfo{author}{Jiang, Z.}, \bibinfo{author}{Duelmer, F.}, \bibinfo{author}{Huang, D.}, \bibinfo{author}{Navab, N.}, \bibinfo{year}{2024}.
\newblock \bibinfo{title}{Machine learning in robotic ultrasound imaging: Challenges and perspectives}.
\newblock \bibinfo{journal}{Annual Review of Control, Robotics, and Autonomous Systems} \bibinfo{volume}{7}.
\bibitem[{Bilic et~al.(2023)Bilic, Christ, Li, Vorontsov, Ben-Cohen, Kaissis, Szeskin, Jacobs, Mamani, Chartrand et~al.}]{bilic2023liver}
\bibinfo{author}{Bilic, P.}, \bibinfo{author}{Christ, P.}, \bibinfo{author}{Li, H.B.}, \bibinfo{author}{Vorontsov, E.}, \bibinfo{author}{Ben-Cohen, A.}, \bibinfo{author}{Kaissis, G.}, \bibinfo{author}{Szeskin, A.}, \bibinfo{author}{Jacobs, C.}, \bibinfo{author}{Mamani, G.E.H.}, \bibinfo{author}{Chartrand, G.}, et~al., \bibinfo{year}{2023}.
\newblock \bibinfo{title}{The liver tumor segmentation benchmark (lits)}.
\newblock \bibinfo{journal}{Medical Image Analysis} \bibinfo{volume}{84}, \bibinfo{pages}{102680}.
\bibitem[{Bond-Taylor et~al.(2021)Bond-Taylor, Leach, Long and Willcocks}]{bond2021deep}
\bibinfo{author}{Bond-Taylor, S.}, \bibinfo{author}{Leach, A.}, \bibinfo{author}{Long, Y.}, \bibinfo{author}{Willcocks, C.G.}, \bibinfo{year}{2021}.
\newblock \bibinfo{title}{Deep generative modelling: A comparative review of vaes, gans, normalizing flows, energy-based and autoregressive models}.
\newblock \bibinfo{journal}{IEEE transactions on pattern analysis and machine intelligence} \bibinfo{volume}{44}, \bibinfo{pages}{7327--7347}.
\bibitem[{Cai et~al.(2023)Cai, Chen, Yang, Zhou and Cheng}]{cai2023dual}
\bibinfo{author}{Cai, Y.}, \bibinfo{author}{Chen, H.}, \bibinfo{author}{Yang, X.}, \bibinfo{author}{Zhou, Y.}, \bibinfo{author}{Cheng, K.T.}, \bibinfo{year}{2023}.
\newblock \bibinfo{title}{Dual-distribution discrepancy with self-supervised refinement for anomaly detection in medical images}.
\newblock \bibinfo{journal}{Medical image analysis} \bibinfo{volume}{86}, \bibinfo{pages}{102794}.
\bibitem[{Chen et~al.(2020)Chen, You, Tezcan and Konukoglu}]{chen2020unsupervised}
\bibinfo{author}{Chen, X.}, \bibinfo{author}{You, S.}, \bibinfo{author}{Tezcan, K.C.}, \bibinfo{author}{Konukoglu, E.}, \bibinfo{year}{2020}.
\newblock \bibinfo{title}{Unsupervised lesion detection via image restoration with a normative prior}.
\newblock \bibinfo{journal}{Medical image analysis} \bibinfo{volume}{64}, \bibinfo{pages}{101713}.
\bibitem[{Dhariwal and Nichol(2021)}]{dhariwal2021diffusion}
\bibinfo{author}{Dhariwal, P.}, \bibinfo{author}{Nichol, A.}, \bibinfo{year}{2021}.
\newblock \bibinfo{title}{Diffusion models beat gans on image synthesis}.
\newblock \bibinfo{journal}{Advances in neural information processing systems} \bibinfo{volume}{34}, \bibinfo{pages}{8780--8794}.
\bibitem[{Dom{\'\i}nguez et~al.(2024)Dom{\'\i}nguez, Velikova, Navab and Azampour}]{dominguez2024diffusion}
\bibinfo{author}{Dom{\'\i}nguez, M.}, \bibinfo{author}{Velikova, Y.}, \bibinfo{author}{Navab, N.}, \bibinfo{author}{Azampour, M.F.}, \bibinfo{year}{2024}.
\newblock \bibinfo{title}{Diffusion as sound propagation: Physics-inspired model for ultrasound image generation}, in: \bibinfo{booktitle}{International Conference on Medical Image Computing and Computer-Assisted Intervention}, \bibinfo{organization}{Springer}. pp. \bibinfo{pages}{613--623}.
\bibitem[{Du et~al.(2023)Du, Jiang, Tan, Wu, Dou, Li, Li and Wan}]{du2023arsdm}
\bibinfo{author}{Du, Y.}, \bibinfo{author}{Jiang, Y.}, \bibinfo{author}{Tan, S.}, \bibinfo{author}{Wu, X.}, \bibinfo{author}{Dou, Q.}, \bibinfo{author}{Li, Z.}, \bibinfo{author}{Li, G.}, \bibinfo{author}{Wan, X.}, \bibinfo{year}{2023}.
\newblock \bibinfo{title}{Arsdm: colonoscopy images synthesis with adaptive refinement semantic diffusion models}, in: \bibinfo{booktitle}{International conference on medical image computing and computer-assisted intervention}, \bibinfo{organization}{Springer}. pp. \bibinfo{pages}{339--349}.
\bibitem[{Duque et~al.(2024)Duque, Marquardt, Velikova, Lacourpaille, Nordez, Crouzier, Lee, Mateus and Navab}]{duque2024ultrasound}
\bibinfo{author}{Duque, V.G.}, \bibinfo{author}{Marquardt, A.}, \bibinfo{author}{Velikova, Y.}, \bibinfo{author}{Lacourpaille, L.}, \bibinfo{author}{Nordez, A.}, \bibinfo{author}{Crouzier, M.}, \bibinfo{author}{Lee, H.J.}, \bibinfo{author}{Mateus, D.}, \bibinfo{author}{Navab, N.}, \bibinfo{year}{2024}.
\newblock \bibinfo{title}{Ultrasound segmentation analysis via distinct and completed anatomical borders}.
\newblock \bibinfo{journal}{International Journal of Computer Assisted Radiology and Surgery} , \bibinfo{pages}{1--9}.
\bibitem[{Fernando et~al.(2021)Fernando, Gammulle, Denman, Sridharan and Fookes}]{fernando2021deep}
\bibinfo{author}{Fernando, T.}, \bibinfo{author}{Gammulle, H.}, \bibinfo{author}{Denman, S.}, \bibinfo{author}{Sridharan, S.}, \bibinfo{author}{Fookes, C.}, \bibinfo{year}{2021}.
\newblock \bibinfo{title}{Deep learning for medical anomaly detection--a survey}.
\newblock \bibinfo{journal}{ACM Computing Surveys (CSUR)} \bibinfo{volume}{54}, \bibinfo{pages}{1--37}.
\bibitem[{Fontanella et~al.(2023)Fontanella, Antoniou, Li, Wardlaw, Mair, Trucco and Storkey}]{fontanella2023acat}
\bibinfo{author}{Fontanella, A.}, \bibinfo{author}{Antoniou, A.}, \bibinfo{author}{Li, W.}, \bibinfo{author}{Wardlaw, J.}, \bibinfo{author}{Mair, G.}, \bibinfo{author}{Trucco, E.}, \bibinfo{author}{Storkey, A.}, \bibinfo{year}{2023}.
\newblock \bibinfo{title}{{ACAT}: Adversarial counterfactual attention for classification and detection in medical imaging}, in: \bibinfo{booktitle}{Proceedings of the 40th International Conference on Machine Learning}, \bibinfo{publisher}{PMLR}. pp. \bibinfo{pages}{10153--10169}.
\bibitem[{Fontanella et~al.(2024)Fontanella, Mair, Wardlaw, Trucco and Storkey}]{fontanella2024diffusion}
\bibinfo{author}{Fontanella, A.}, \bibinfo{author}{Mair, G.}, \bibinfo{author}{Wardlaw, J.}, \bibinfo{author}{Trucco, E.}, \bibinfo{author}{Storkey, A.}, \bibinfo{year}{2024}.
\newblock \bibinfo{title}{Diffusion models for counterfactual generation and anomaly detection in brain images}.
\newblock \bibinfo{journal}{IEEE Transactions on Medical Imaging} .
\bibitem[{Frotscher et~al.(2023)Frotscher, Kapoor, Wolfers and Baumgartner}]{frotscher2023unsupervised}
\bibinfo{author}{Frotscher, A.}, \bibinfo{author}{Kapoor, J.}, \bibinfo{author}{Wolfers, T.}, \bibinfo{author}{Baumgartner, C.F.}, \bibinfo{year}{2023}.
\newblock \bibinfo{title}{Unsupervised anomaly detection using aggregated normative diffusion}.
\newblock \bibinfo{journal}{arXiv preprint arXiv:2312.01904} .
\bibitem[{Gong et~al.(2019)Gong, Liu, Le, Saha, Mansour, Venkatesh and Hengel}]{gong2019memorizing}
\bibinfo{author}{Gong, D.}, \bibinfo{author}{Liu, L.}, \bibinfo{author}{Le, V.}, \bibinfo{author}{Saha, B.}, \bibinfo{author}{Mansour, M.R.}, \bibinfo{author}{Venkatesh, S.}, \bibinfo{author}{Hengel, A.v.d.}, \bibinfo{year}{2019}.
\newblock \bibinfo{title}{Memorizing normality to detect anomaly: Memory-augmented deep autoencoder for unsupervised anomaly detection}, in: \bibinfo{booktitle}{Proceedings of the IEEE/CVF international conference on computer vision}, pp. \bibinfo{pages}{1705--1714}.
\bibitem[{Ho et~al.(2020)Ho, Jain and Abbeel}]{ho2020denoising}
\bibinfo{author}{Ho, J.}, \bibinfo{author}{Jain, A.}, \bibinfo{author}{Abbeel, P.}, \bibinfo{year}{2020}.
\newblock \bibinfo{title}{Denoising diffusion probabilistic models}.
\newblock \bibinfo{journal}{Advances in neural information processing systems} \bibinfo{volume}{33}, \bibinfo{pages}{6840--6851}.
\bibitem[{Huang et~al.(2025)Huang, Li, Karlas, Chu, Au, Navab and Jiang}]{huang2025vibnet}
\bibinfo{author}{Huang, D.}, \bibinfo{author}{Li, C.}, \bibinfo{author}{Karlas, A.}, \bibinfo{author}{Chu, X.}, \bibinfo{author}{Au, K.S.}, \bibinfo{author}{Navab, N.}, \bibinfo{author}{Jiang, Z.}, \bibinfo{year}{2025}.
\newblock \bibinfo{title}{Vibnet: Vibration-boosted needle detection in ultrasound images}.
\newblock \bibinfo{journal}{IEEE Transactions on Medical Imaging} .
\bibitem[{Jiang et~al.(2024a)Jiang, Zhang, Wen, Cui, Lu, Rekik, Ma and Chen}]{jiang2024self}
\bibinfo{author}{Jiang, H.}, \bibinfo{author}{Zhang, S.}, \bibinfo{author}{Wen, X.}, \bibinfo{author}{Cui, H.}, \bibinfo{author}{Lu, J.}, \bibinfo{author}{Rekik, I.}, \bibinfo{author}{Ma, J.}, \bibinfo{author}{Chen, G.}, \bibinfo{year}{2024}a.
\newblock \bibinfo{title}{Self-supervised denoising of diffusion mri data via spatio-angular noise2noise}, in: \bibinfo{booktitle}{2024 IEEE International Symposium on Biomedical Imaging (ISBI)}, \bibinfo{organization}{IEEE}. pp. \bibinfo{pages}{1--5}.
\bibitem[{Jiang et~al.(2024b)Jiang, Bi, Zhou, Hu, Burke and Navab}]{jiang2024intelligent}
\bibinfo{author}{Jiang, Z.}, \bibinfo{author}{Bi, Y.}, \bibinfo{author}{Zhou, M.}, \bibinfo{author}{Hu, Y.}, \bibinfo{author}{Burke, M.}, \bibinfo{author}{Navab, N.}, \bibinfo{year}{2024}b.
\newblock \bibinfo{title}{Intelligent robotic sonographer: Mutual information-based disentangled reward learning from few demonstrations}.
\newblock \bibinfo{journal}{The International Journal of Robotics Research} \bibinfo{volume}{43}, \bibinfo{pages}{981--1002}.
\bibitem[{Jiang et~al.(2024c)Jiang, Kang, Bi, Li, Li and Navab}]{jiang2024class}
\bibinfo{author}{Jiang, Z.}, \bibinfo{author}{Kang, Y.}, \bibinfo{author}{Bi, Y.}, \bibinfo{author}{Li, X.}, \bibinfo{author}{Li, C.}, \bibinfo{author}{Navab, N.}, \bibinfo{year}{2024}c.
\newblock \bibinfo{title}{Class-aware cartilage segmentation for autonomous us-ct registration in robotic intercostal ultrasound imaging}.
\newblock \bibinfo{journal}{IEEE Transactions on Automation Science and Engineering} .
\bibitem[{Jiang et~al.(2023a)Jiang, Salcudean and Navab}]{jiang2023robotic}
\bibinfo{author}{Jiang, Z.}, \bibinfo{author}{Salcudean, S.E.}, \bibinfo{author}{Navab, N.}, \bibinfo{year}{2023}a.
\newblock \bibinfo{title}{Robotic ultrasound imaging: State-of-the-art and future perspectives}.
\newblock \bibinfo{journal}{Medical image analysis} \bibinfo{volume}{89}, \bibinfo{pages}{102878}.
\bibitem[{Jiang et~al.(2023b)Jiang, Zhou, Cao and Navab}]{jiang2023defcor}
\bibinfo{author}{Jiang, Z.}, \bibinfo{author}{Zhou, Y.}, \bibinfo{author}{Cao, D.}, \bibinfo{author}{Navab, N.}, \bibinfo{year}{2023}b.
\newblock \bibinfo{title}{Defcor-net: Physics-aware ultrasound deformation correction}.
\newblock \bibinfo{journal}{Medical Image Analysis} \bibinfo{volume}{90}, \bibinfo{pages}{102923}.
\bibitem[{Kascenas et~al.(2022)Kascenas, Pugeault and O’Neil}]{kascenas2022denoising}
\bibinfo{author}{Kascenas, A.}, \bibinfo{author}{Pugeault, N.}, \bibinfo{author}{O’Neil, A.Q.}, \bibinfo{year}{2022}.
\newblock \bibinfo{title}{Denoising autoencoders for unsupervised anomaly detection in brain mri}, in: \bibinfo{booktitle}{International Conference on Medical Imaging with Deep Learning}, \bibinfo{organization}{PMLR}. pp. \bibinfo{pages}{653--664}.
\bibitem[{Kascenas et~al.(2023)Kascenas, Sanchez, Schrempf, Wang, Clackett, Mikhael, Voisey, Goatman, Weir, Pugeault et~al.}]{kascenas2023role}
\bibinfo{author}{Kascenas, A.}, \bibinfo{author}{Sanchez, P.}, \bibinfo{author}{Schrempf, P.}, \bibinfo{author}{Wang, C.}, \bibinfo{author}{Clackett, W.}, \bibinfo{author}{Mikhael, S.S.}, \bibinfo{author}{Voisey, J.P.}, \bibinfo{author}{Goatman, K.}, \bibinfo{author}{Weir, A.}, \bibinfo{author}{Pugeault, N.}, et~al., \bibinfo{year}{2023}.
\newblock \bibinfo{title}{The role of noise in denoising models for anomaly detection in medical images}.
\newblock \bibinfo{journal}{Medical Image Analysis} \bibinfo{volume}{90}, \bibinfo{pages}{102963}.
\bibitem[{Kazerouni et~al.(2023)Kazerouni, Aghdam, Heidari, Azad, Fayyaz, Hacihaliloglu and Merhof}]{kazerouni2023diffusion}
\bibinfo{author}{Kazerouni, A.}, \bibinfo{author}{Aghdam, E.K.}, \bibinfo{author}{Heidari, M.}, \bibinfo{author}{Azad, R.}, \bibinfo{author}{Fayyaz, M.}, \bibinfo{author}{Hacihaliloglu, I.}, \bibinfo{author}{Merhof, D.}, \bibinfo{year}{2023}.
\newblock \bibinfo{title}{Diffusion models in medical imaging: A comprehensive survey}.
\newblock \bibinfo{journal}{Medical Image Analysis} \bibinfo{volume}{88}, \bibinfo{pages}{102846}.
\bibitem[{Kingma et~al.(2013)Kingma, Welling et~al.}]{kingma2013auto}
\bibinfo{author}{Kingma, D.P.}, \bibinfo{author}{Welling, M.}, et~al., \bibinfo{year}{2013}.
\newblock \bibinfo{title}{Auto-encoding variational bayes}.
\bibitem[{Li et~al.(2021)Li, Sohn, Yoon and Pfister}]{li2021cutpaste}
\bibinfo{author}{Li, C.L.}, \bibinfo{author}{Sohn, K.}, \bibinfo{author}{Yoon, J.}, \bibinfo{author}{Pfister, T.}, \bibinfo{year}{2021}.
\newblock \bibinfo{title}{Cutpaste: Self-supervised learning for anomaly detection and localization}, in: \bibinfo{booktitle}{Proceedings of the IEEE/CVF conference on computer vision and pattern recognition}, pp. \bibinfo{pages}{9664--9674}.
\bibitem[{Li et~al.(2023a)Li, Cao, Wang, Liu, Dou, Chen and Heng}]{li2023fast}
\bibinfo{author}{Li, J.}, \bibinfo{author}{Cao, H.}, \bibinfo{author}{Wang, J.}, \bibinfo{author}{Liu, F.}, \bibinfo{author}{Dou, Q.}, \bibinfo{author}{Chen, G.}, \bibinfo{author}{Heng, P.A.}, \bibinfo{year}{2023}a.
\newblock \bibinfo{title}{Fast non-markovian diffusion model for weakly supervised anomaly detection in brain mr images}, in: \bibinfo{booktitle}{International Conference on Medical Image Computing and Computer-Assisted Intervention}, \bibinfo{organization}{Springer}. pp. \bibinfo{pages}{579--589}.
\bibitem[{Li et~al.(2023b)Li, Lao, Kang, Jiang, Du, Zhang and Li}]{li2023self}
\bibinfo{author}{Li, Y.}, \bibinfo{author}{Lao, Q.}, \bibinfo{author}{Kang, Q.}, \bibinfo{author}{Jiang, Z.}, \bibinfo{author}{Du, S.}, \bibinfo{author}{Zhang, S.}, \bibinfo{author}{Li, K.}, \bibinfo{year}{2023}b.
\newblock \bibinfo{title}{Self-supervised anomaly detection, staging and segmentation for retinal images}.
\newblock \bibinfo{journal}{Medical Image Analysis} \bibinfo{volume}{87}, \bibinfo{pages}{102805}.
\bibitem[{Liang et~al.(2024)Liang, Guo, Noble and Kamnitsas}]{liang2024itermask2}
\bibinfo{author}{Liang, Z.}, \bibinfo{author}{Guo, X.}, \bibinfo{author}{Noble, J.A.}, \bibinfo{author}{Kamnitsas, K.}, \bibinfo{year}{2024}.
\newblock \bibinfo{title}{Itermask 2: Iterative unsupervised anomaly segmentation via spatial and frequency masking for brain lesions in mri}, in: \bibinfo{booktitle}{International Conference on Medical Image Computing and Computer-Assisted Intervention}, \bibinfo{organization}{Springer}. pp. \bibinfo{pages}{339--348}.
\bibitem[{Mao et~al.(2020)Mao, Xue, Wang, Zhang, Zheng and Liu}]{mao2020abnormality}
\bibinfo{author}{Mao, Y.}, \bibinfo{author}{Xue, F.F.}, \bibinfo{author}{Wang, R.}, \bibinfo{author}{Zhang, J.}, \bibinfo{author}{Zheng, W.S.}, \bibinfo{author}{Liu, H.}, \bibinfo{year}{2020}.
\newblock \bibinfo{title}{Abnormality detection in chest x-ray images using uncertainty prediction autoencoders}, in: \bibinfo{booktitle}{Medical Image Computing and Computer Assisted Intervention--MICCAI 2020: 23rd International Conference, Lima, Peru, October 4--8, 2020, Proceedings, Part VI 23}, \bibinfo{organization}{Springer}. pp. \bibinfo{pages}{529--538}.
\bibitem[{Marimont et~al.(2024)Marimont, Siomos, Baugh, Tzelepis, Kainz and Tarroni}]{marimont2024ensembled}
\bibinfo{author}{Marimont, S.N.}, \bibinfo{author}{Siomos, V.}, \bibinfo{author}{Baugh, M.}, \bibinfo{author}{Tzelepis, C.}, \bibinfo{author}{Kainz, B.}, \bibinfo{author}{Tarroni, G.}, \bibinfo{year}{2024}.
\newblock \bibinfo{title}{Ensembled cold-diffusion restorations for unsupervised anomaly detection}, in: \bibinfo{booktitle}{Proceedings of the 27th International Conference on Medical Image Computing and Computer-Assisted Intervention (MICCAI 2024)}, \bibinfo{publisher}{Springer Nature Switzerland}. pp. \bibinfo{pages}{243--253}.
\bibitem[{Medghalchi et~al.(2025)Medghalchi, Heidari, Allard, Sigal and Hacihaliloglu}]{medghalchi2024prompt2perturb}
\bibinfo{author}{Medghalchi, Y.}, \bibinfo{author}{Heidari, M.}, \bibinfo{author}{Allard, C.}, \bibinfo{author}{Sigal, L.}, \bibinfo{author}{Hacihaliloglu, I.}, \bibinfo{year}{2025}.
\newblock \bibinfo{title}{Prompt2perturb (p2p): Text-guided diffusion-based adversarial attacks on breast ultrasound images}.
\newblock \bibinfo{journal}{CVPR} .
\bibitem[{Menze et~al.(2014)Menze, Jakab, Bauer, Kalpathy-Cramer, Farahani, Kirby, Burren, Porz, Slotboom, Wiest et~al.}]{menze2014multimodal}
\bibinfo{author}{Menze, B.H.}, \bibinfo{author}{Jakab, A.}, \bibinfo{author}{Bauer, S.}, \bibinfo{author}{Kalpathy-Cramer, J.}, \bibinfo{author}{Farahani, K.}, \bibinfo{author}{Kirby, J.}, \bibinfo{author}{Burren, Y.}, \bibinfo{author}{Porz, N.}, \bibinfo{author}{Slotboom, J.}, \bibinfo{author}{Wiest, R.}, et~al., \bibinfo{year}{2014}.
\newblock \bibinfo{title}{The multimodal brain tumor image segmentation benchmark (brats)}.
\newblock \bibinfo{journal}{IEEE transactions on medical imaging} \bibinfo{volume}{34}, \bibinfo{pages}{1993--2024}.
\bibitem[{Mykula et~al.(2024)Mykula, Gasser, Lobmaier, Schnabel, Zimmer and Bercea}]{mykula2024diffusion}
\bibinfo{author}{Mykula, H.}, \bibinfo{author}{Gasser, L.}, \bibinfo{author}{Lobmaier, S.}, \bibinfo{author}{Schnabel, J.A.}, \bibinfo{author}{Zimmer, V.}, \bibinfo{author}{Bercea, C.I.}, \bibinfo{year}{2024}.
\newblock \bibinfo{title}{Diffusion models for unsupervised anomaly detection in fetal brain ultrasound}, in: \bibinfo{booktitle}{International Workshop on Advances in Simplifying Medical Ultrasound}, \bibinfo{organization}{Springer}. pp. \bibinfo{pages}{220--230}.
\bibitem[{Nichol and Dhariwal(2021)}]{nichol2021improved}
\bibinfo{author}{Nichol, A.Q.}, \bibinfo{author}{Dhariwal, P.}, \bibinfo{year}{2021}.
\newblock \bibinfo{title}{Improved denoising diffusion probabilistic models}, in: \bibinfo{booktitle}{International conference on machine learning}, \bibinfo{organization}{PMLR}. pp. \bibinfo{pages}{8162--8171}.
\bibitem[{Olsen et~al.(2024)Olsen, Ambsdorf, Lin, Taks{\o}e-Vester, Svendsen, Christensen, Nielsen, Tolsgaard, Feragen and Pegios}]{olsen2024unsupervised}
\bibinfo{author}{Olsen, M.D.S.}, \bibinfo{author}{Ambsdorf, J.}, \bibinfo{author}{Lin, M.}, \bibinfo{author}{Taks{\o}e-Vester, C.}, \bibinfo{author}{Svendsen, M.B.S.}, \bibinfo{author}{Christensen, A.N.}, \bibinfo{author}{Nielsen, M.}, \bibinfo{author}{Tolsgaard, M.G.}, \bibinfo{author}{Feragen, A.}, \bibinfo{author}{Pegios, P.}, \bibinfo{year}{2024}.
\newblock \bibinfo{title}{Unsupervised detection of fetal brain anomalies using denoising diffusion models}, in: \bibinfo{booktitle}{International Workshop on Advances in Simplifying Medical Ultrasound}, \bibinfo{organization}{Springer}. pp. \bibinfo{pages}{209--219}.
\bibitem[{Pawlowski et~al.(2018)Pawlowski, Lee, Rajchl, McDonagh, Ferrante, Kamnitsas, Cooke, Stevenson, Khetani, Newman et~al.}]{pawlowski2018unsupervised}
\bibinfo{author}{Pawlowski, N.}, \bibinfo{author}{Lee, M.}, \bibinfo{author}{Rajchl, M.}, \bibinfo{author}{McDonagh, S.}, \bibinfo{author}{Ferrante, E.}, \bibinfo{author}{Kamnitsas, K.}, \bibinfo{author}{Cooke, S.}, \bibinfo{author}{Stevenson, S.}, \bibinfo{author}{Khetani, A.}, \bibinfo{author}{Newman, T.}, et~al., \bibinfo{year}{2018}.
\newblock \bibinfo{title}{Unsupervised lesion detection in brain ct using bayesian convolutional autoencoders}.
\newblock \bibinfo{journal}{Medical Imaging with Deep Learning} , \bibinfo{pages}{17}.
\bibitem[{Pinaya et~al.(2022)Pinaya, Graham, Gray, Da~Costa, Tudosiu, Wright, Mah, MacKinnon, Teo, Jager et~al.}]{pinaya2022fast}
\bibinfo{author}{Pinaya, W.H.}, \bibinfo{author}{Graham, M.S.}, \bibinfo{author}{Gray, R.}, \bibinfo{author}{Da~Costa, P.F.}, \bibinfo{author}{Tudosiu, P.D.}, \bibinfo{author}{Wright, P.}, \bibinfo{author}{Mah, Y.H.}, \bibinfo{author}{MacKinnon, A.D.}, \bibinfo{author}{Teo, J.T.}, \bibinfo{author}{Jager, R.}, et~al., \bibinfo{year}{2022}.
\newblock \bibinfo{title}{Fast unsupervised brain anomaly detection and segmentation with diffusion models}, in: \bibinfo{booktitle}{International Conference on Medical Image Computing and Computer-Assisted Intervention}, \bibinfo{organization}{Springer}. pp. \bibinfo{pages}{705--714}.
\bibitem[{Ravi et~al.(2025)Ravi, Gabeur, Hu, Hu, Ryali, Ma, Khedr, R{\"a}dle, Rolland, Gustafson, Mintun, Pan, Alwala, Carion, Wu, Girshick, Dollar and Feichtenhofer}]{ravi2025sam}
\bibinfo{author}{Ravi, N.}, \bibinfo{author}{Gabeur, V.}, \bibinfo{author}{Hu, Y.T.}, \bibinfo{author}{Hu, R.}, \bibinfo{author}{Ryali, C.}, \bibinfo{author}{Ma, T.}, \bibinfo{author}{Khedr, H.}, \bibinfo{author}{R{\"a}dle, R.}, \bibinfo{author}{Rolland, C.}, \bibinfo{author}{Gustafson, L.}, \bibinfo{author}{Mintun, E.}, \bibinfo{author}{Pan, J.}, \bibinfo{author}{Alwala, K.V.}, \bibinfo{author}{Carion, N.}, \bibinfo{author}{Wu, C.Y.}, \bibinfo{author}{Girshick, R.}, \bibinfo{author}{Dollar, P.}, \bibinfo{author}{Feichtenhofer, C.}, \bibinfo{year}{2025}.
\newblock \bibinfo{title}{{SAM} 2: Segment anything in images and videos}, in: \bibinfo{booktitle}{Proceedings of the Thirteenth International Conference on Learning Representations (ICLR)}.
\bibitem[{Ronneberger et~al.(2015)Ronneberger, Fischer and Brox}]{ronneberger2015u}
\bibinfo{author}{Ronneberger, O.}, \bibinfo{author}{Fischer, P.}, \bibinfo{author}{Brox, T.}, \bibinfo{year}{2015}.
\newblock \bibinfo{title}{U-net: Convolutional networks for biomedical image segmentation}, in: \bibinfo{booktitle}{Medical image computing and computer-assisted intervention--MICCAI 2015: 18th international conference, Munich, Germany, October 5-9, 2015, proceedings, part III 18}, \bibinfo{organization}{Springer}. pp. \bibinfo{pages}{234--241}.
\bibitem[{Schlegl et~al.(2019)Schlegl, Seeb{\"o}ck, Waldstein, Langs and Schmidt-Erfurth}]{schlegl2019f}
\bibinfo{author}{Schlegl, T.}, \bibinfo{author}{Seeb{\"o}ck, P.}, \bibinfo{author}{Waldstein, S.M.}, \bibinfo{author}{Langs, G.}, \bibinfo{author}{Schmidt-Erfurth, U.}, \bibinfo{year}{2019}.
\newblock \bibinfo{title}{f-anogan: Fast unsupervised anomaly detection with generative adversarial networks}.
\newblock \bibinfo{journal}{Medical image analysis} \bibinfo{volume}{54}, \bibinfo{pages}{30--44}.
\bibitem[{Schlegl et~al.(2017)Schlegl, Seeb{\"o}ck, Waldstein, Schmidt-Erfurth and Langs}]{schlegl2017unsupervised}
\bibinfo{author}{Schlegl, T.}, \bibinfo{author}{Seeb{\"o}ck, P.}, \bibinfo{author}{Waldstein, S.M.}, \bibinfo{author}{Schmidt-Erfurth, U.}, \bibinfo{author}{Langs, G.}, \bibinfo{year}{2017}.
\newblock \bibinfo{title}{Unsupervised anomaly detection with generative adversarial networks to guide marker discovery}, in: \bibinfo{booktitle}{International conference on information processing in medical imaging}, \bibinfo{organization}{Springer}. pp. \bibinfo{pages}{146--157}.
\bibitem[{Schl{\"u}ter et~al.(2022)Schl{\"u}ter, Tan, Hou and Kainz}]{schluter2022natural}
\bibinfo{author}{Schl{\"u}ter, H.M.}, \bibinfo{author}{Tan, J.}, \bibinfo{author}{Hou, B.}, \bibinfo{author}{Kainz, B.}, \bibinfo{year}{2022}.
\newblock \bibinfo{title}{Natural synthetic anomalies for self-supervised anomaly detection and localization}, in: \bibinfo{booktitle}{European Conference on Computer Vision}, \bibinfo{organization}{Springer}. pp. \bibinfo{pages}{474--489}.
\bibitem[{Siddiquee et~al.(2024)Siddiquee, Shah, Wu, Chong, Schwedt, Dumkrieger, Nikolova and Li}]{siddiquee2024brainomaly}
\bibinfo{author}{Siddiquee, M.M.R.}, \bibinfo{author}{Shah, J.}, \bibinfo{author}{Wu, T.}, \bibinfo{author}{Chong, C.}, \bibinfo{author}{Schwedt, T.J.}, \bibinfo{author}{Dumkrieger, G.}, \bibinfo{author}{Nikolova, S.}, \bibinfo{author}{Li, B.}, \bibinfo{year}{2024}.
\newblock \bibinfo{title}{Brainomaly: Unsupervised neurologic disease detection utilizing unannotated t1-weighted brain mr images}, in: \bibinfo{booktitle}{Proceedings of the IEEE/CVF Winter Conference on Applications of Computer Vision}, pp. \bibinfo{pages}{7573--7582}.
\bibitem[{Song et~al.(2021)Song, Meng and Ermon}]{song2020denoising}
\bibinfo{author}{Song, J.}, \bibinfo{author}{Meng, C.}, \bibinfo{author}{Ermon, S.}, \bibinfo{year}{2021}.
\newblock \bibinfo{title}{Denoising diffusion implicit models}, in: \bibinfo{booktitle}{International Conference on Learning Representations}.
\bibitem[{Tan et~al.(2022)Tan, Hou, Batten, Qiu and Kainz}]{tan2020detecting}
\bibinfo{author}{Tan, J.}, \bibinfo{author}{Hou, B.}, \bibinfo{author}{Batten, J.}, \bibinfo{author}{Qiu, H.}, \bibinfo{author}{Kainz, B.}, \bibinfo{year}{2022}.
\newblock \bibinfo{title}{Detecting outliers with foreign patch interpolation}.
\newblock \bibinfo{journal}{Machine Learning for Biomedical Imaging} \bibinfo{volume}{1}, \bibinfo{pages}{1--27}.
\bibitem[{Tan et~al.(2021)Tan, Hou, Day, Simpson, Rueckert and Kainz}]{tan2021detecting}
\bibinfo{author}{Tan, J.}, \bibinfo{author}{Hou, B.}, \bibinfo{author}{Day, T.}, \bibinfo{author}{Simpson, J.}, \bibinfo{author}{Rueckert, D.}, \bibinfo{author}{Kainz, B.}, \bibinfo{year}{2021}.
\newblock \bibinfo{title}{Detecting outliers with poisson image interpolation}, in: \bibinfo{booktitle}{Medical Image Computing and Computer Assisted Intervention--MICCAI 2021: 24th International Conference, Strasbourg, France, September 27--October 1, 2021, Proceedings, Part V 24}, \bibinfo{organization}{Springer}. pp. \bibinfo{pages}{581--591}.
\bibitem[{Uzunova et~al.(2019)Uzunova, Schultz, Handels and Ehrhardt}]{uzunova2019unsupervised}
\bibinfo{author}{Uzunova, H.}, \bibinfo{author}{Schultz, S.}, \bibinfo{author}{Handels, H.}, \bibinfo{author}{Ehrhardt, J.}, \bibinfo{year}{2019}.
\newblock \bibinfo{title}{Unsupervised pathology detection in medical images using conditional variational autoencoders}.
\newblock \bibinfo{journal}{International journal of computer assisted radiology and surgery} \bibinfo{volume}{14}, \bibinfo{pages}{451--461}.
\bibitem[{Wolleb et~al.(2022)Wolleb, Bieder, Sandk{\"u}hler and Cattin}]{wolleb2022diffusion}
\bibinfo{author}{Wolleb, J.}, \bibinfo{author}{Bieder, F.}, \bibinfo{author}{Sandk{\"u}hler, R.}, \bibinfo{author}{Cattin, P.C.}, \bibinfo{year}{2022}.
\newblock \bibinfo{title}{Diffusion models for medical anomaly detection}, in: \bibinfo{booktitle}{International Conference on Medical image computing and computer-assisted intervention}, \bibinfo{organization}{Springer}. pp. \bibinfo{pages}{35--45}.
\bibitem[{Wyatt et~al.(2022)Wyatt, Leach, Schmon and Willcocks}]{wyatt2022anoddpm}
\bibinfo{author}{Wyatt, J.}, \bibinfo{author}{Leach, A.}, \bibinfo{author}{Schmon, S.M.}, \bibinfo{author}{Willcocks, C.G.}, \bibinfo{year}{2022}.
\newblock \bibinfo{title}{Anoddpm: Anomaly detection with denoising diffusion probabilistic models using simplex noise}, in: \bibinfo{booktitle}{Proceedings of the IEEE/CVF Conference on Computer Vision and Pattern Recognition}, pp. \bibinfo{pages}{650--656}.
\bibitem[{Xu et~al.(2024)Xu, Xu and Giannarou}]{xu2024stereodiffusion}
\bibinfo{author}{Xu, H.}, \bibinfo{author}{Xu, C.}, \bibinfo{author}{Giannarou, S.}, \bibinfo{year}{2024}.
\newblock \bibinfo{title}{Stereodiffusion: Temporally consistent stereo depth estimation with diffusion models}, in: \bibinfo{booktitle}{International Conference on Medical Image Computing and Computer-Assisted Intervention}, \bibinfo{organization}{Springer}. pp. \bibinfo{pages}{596--605}.
\bibitem[{Zhou et~al.(2025)Zhou, Bi, Tong, Wang, Navab and Jiang}]{zhou2025ultraad}
\bibinfo{author}{Zhou, Y.}, \bibinfo{author}{Bi, Y.}, \bibinfo{author}{Tong, W.}, \bibinfo{author}{Wang, W.}, \bibinfo{author}{Navab, N.}, \bibinfo{author}{Jiang, Z.}, \bibinfo{year}{2025}.
\newblock \bibinfo{title}{Ultraad: Fine-grained ultrasound anomaly classification via few-shot clip adaptation}.
\newblock \bibinfo{journal}{arXiv preprint arXiv:2506.19694} .
\bibitem[{Zimmerer et~al.(2018)Zimmerer, Kohl, Petersen, Isensee and Maier-Hein}]{zimmerer2018context}
\bibinfo{author}{Zimmerer, D.}, \bibinfo{author}{Kohl, S.A.}, \bibinfo{author}{Petersen, J.}, \bibinfo{author}{Isensee, F.}, \bibinfo{author}{Maier-Hein, K.H.}, \bibinfo{year}{2018}.
\newblock \bibinfo{title}{Context-encoding variational autoencoder for unsupervised anomaly detection}.
\newblock \bibinfo{journal}{arXiv preprint arXiv:1812.05941} .

\end{thebibliography}



\end{document}